\documentclass[conference]{IEEEtran}
\IEEEoverridecommandlockouts
\usepackage{amsmath,amssymb,amsfonts}
\usepackage{graphicx}
\usepackage{textcomp}
\usepackage{xcolor}
\usepackage{array}
\usepackage{tabularx}
\usepackage{subfigure}
\usepackage{diagbox}
\usepackage{multirow}
\usepackage{url}
\usepackage{balance}
\usepackage{framed}

\newcommand{\RS}[2]{%
\begin{framed}%
\filbreak
\noindent\textbf{Result {#1}:~}{\emph {#2}}%
\end{framed}
}
\def\BibTeX{{\rm B\kern-.05em{\sc i\kern-.025em b}\kern-.08em
    T\kern-.1667em\lower.7ex\hbox{E}\kern-.125emX}}

\begin{document}
\title{Plot2API: Recommending Graphic API from Plot \\via Semantic Parsing Guided Neural Network \\
}

\author{\IEEEauthorblockN{Zeyu Wang$^{1,2}$, Sheng Huang$^{1,2}$\IEEEauthorrefmark{1}, Zhongxin Liu$^{3}$, Meng Yan$^{1,2}$\IEEEauthorrefmark{1}\IEEEauthorrefmark{2}, Xin Xia$^{4}$, Bei Wang$^{2}$, Dan Yang$^{2}$\thanks{\IEEEauthorrefmark{1}Corresponding authors.}\thanks{\IEEEauthorrefmark{2}also with Pengcheng Laboratory, Shenzhen, China.}}
   \IEEEauthorblockA{$^{1}$Key Laboratory of Dependable Service Computing in Cyber Physical Society (Chongqing University),\\ Ministry of Education, China}
   \IEEEauthorblockA{$^{2}$School of Big Data and Software Engineering, Chongqing University, Chongqing, China}
   \IEEEauthorblockA{$^{3}$College of Computer Science and Technology, Zhejiang University, Hangzhou, China}
   \IEEEauthorblockA{$^{4}$Faculty of Information Technology, Monash University, Australia}
   \IEEEauthorblockA{Email:\{zeyuwang, huangsheng, mengy, bwang2013, dyang\}@cqu.edu.cn, liu\_zx@zju.edu.cn, xin.xia@monash.edu}
 }

\maketitle

\begin{abstract}
Plot-based Graphic API recommendation (Plot2API) is an unstudied but meaningful issue, which has several important applications in the context of software engineering and data visualization, such as the plotting guidance of the beginner, graphic API correlation analysis, and code conversion for plotting. Plot2API is a very challenging task, since each plot is often associated with multiple APIs and the appearances of the graphics drawn by the same API can be extremely varied due to the different settings of the parameters. Additionally, the samples of different APIs also suffer from extremely imbalanced.

Considering the lack of technologies in Plot2API, we present a novel deep multi-task learning approach named Semantic Parsing Guided Neural Network (SPGNN) which translates the Plot2API issue as a multi-label image classification and an image semantic parsing tasks for the solution. In SPGNN, the recently advanced Convolutional Neural Network (CNN) named EfficientNet is employed as the backbone network for API recommendation. Meanwhile, a semantic parsing module is complemented to exploit the semantic relevant visual information in feature learning and eliminate the appearance-relevant visual information which may confuse the visual-information-based API recommendation. Moreover, the recent data augmentation technique named random erasing is also applied for alleviating the imbalance of API categories.

We collect plots with the graphic APIs used to drawn them from Stack Overflow, and release three new Plot2API datasets corresponding to the graphic APIs of R and Python programming languages for evaluating the effectiveness of Plot2API techniques. Extensive experimental results not only demonstrate the superiority of our method over the recent deep learning baselines but also show the practicability of our method in the recommendation of graphic APIs.

%

\end{abstract}


\section{Introduction}

\begin{figure}[h]
\centering
\includegraphics[scale = 0.37]{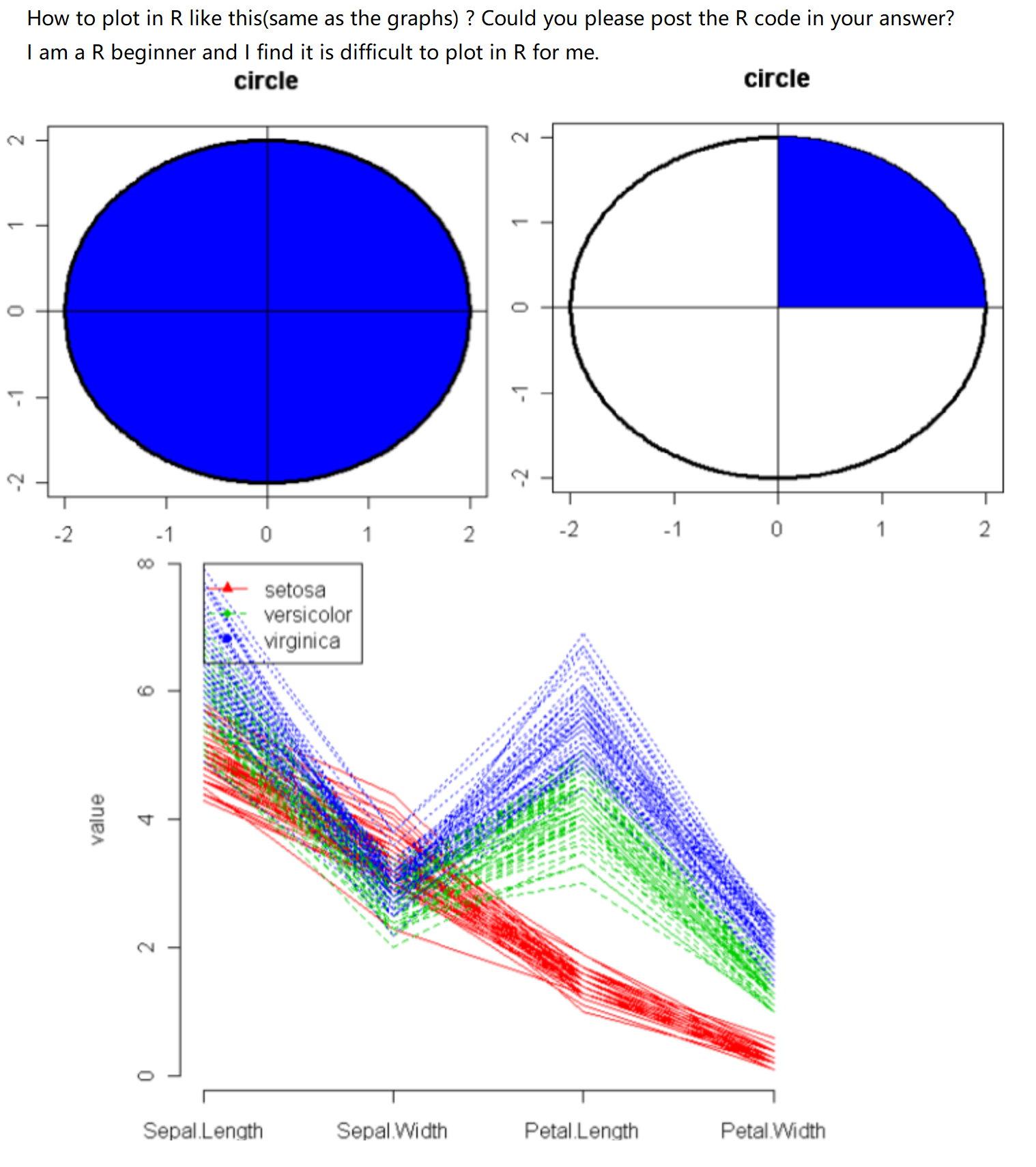}
\vspace{-0.3cm}
\caption{ The help post about Plot2API in Stack Overflow from link: \\ https://stackoverflow.com/questions/12786334/how-to-plot-in-r-like-this}
\label{usercase1}
\vspace{-0.85cm}
\end{figure}

Figures and plots are the indispensable tools for data visualization which provide people with intuitive understanding of data and interaction with data. In software engineering, almost all the programming languages support such functions and possess a series of relevant APIs as one of core libraries or packages. It is very common for the software developer particularly the beginner to search API on the web based on a case figure for guiding the plot. Figure~\ref{usercase1} shows a help post where a developer asks how to draw a figure like the one posts in Stack Overflow. What's more, people might want to know the APIs starting from a plot, such as imitating visualization styles. In agile development, developers often sufficiently utilize the materials of previous projects for speeding up the development, thereby they expect to convert the figures plotted in one language into APIs of the other directly to reduce the time cost. In these scenarios, a tool that can automatically recommend graphic APIs based on a plot can provide guidance for developers and improve their productivity. Therefore, how to identify API based on Plot (Plot2API) is a meaningful task in software engineering and data visualization.

\begin{figure}[t]
\centering
\includegraphics[scale = 0.42]{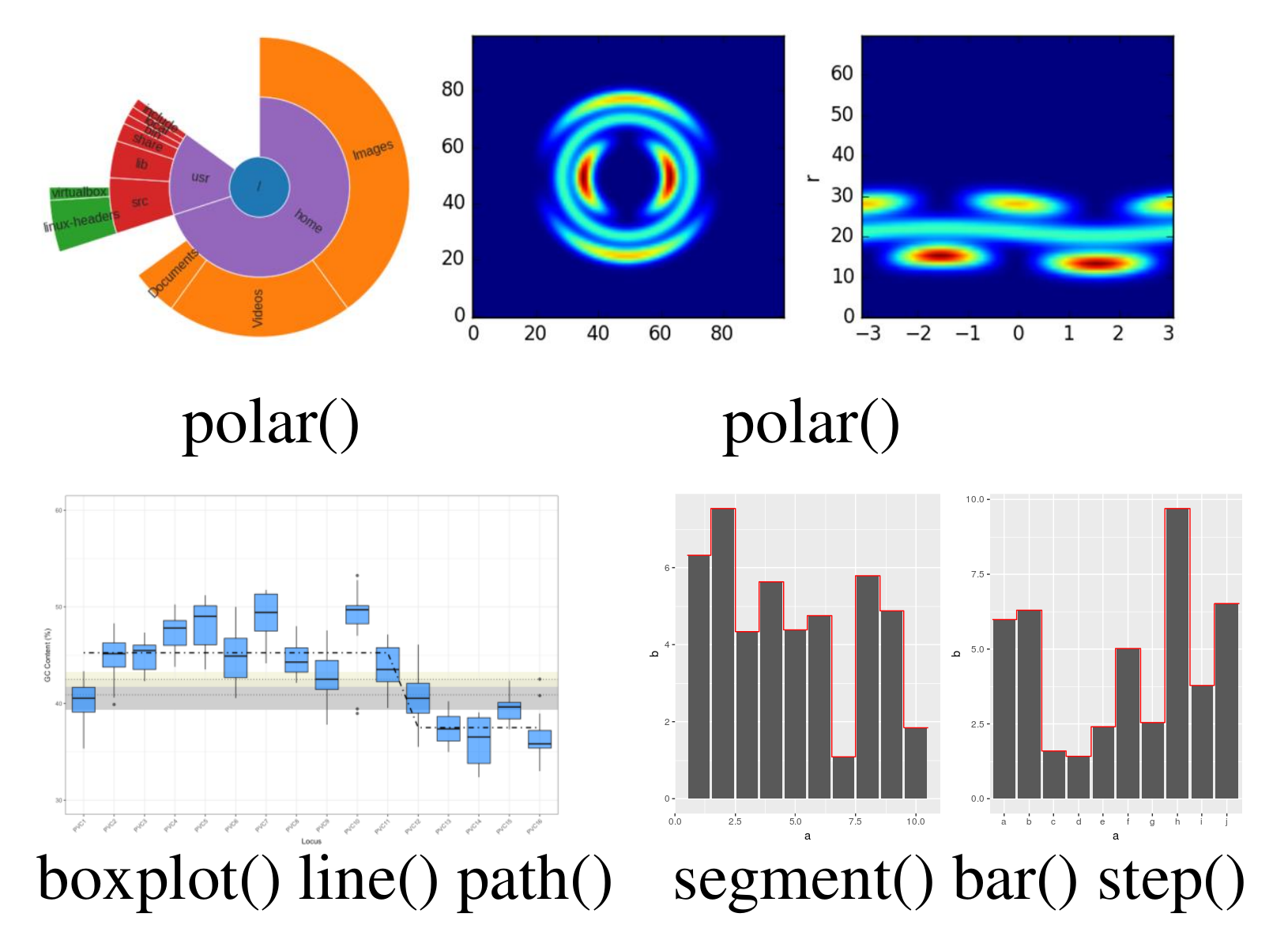}
\vspace{-0.3cm}
\caption{ Examples of data graphics with APIs.}
\label{examples}
\vspace{-0.85cm}
\end{figure}

Plot2API can be deemed as a plot-based API recommendation task, since the set of APIs regards to a programming language is fixed and a plot is often drawn by multiple APIs. API recommendation is not a new issue now in software engineering and many researchers have worked in this direction~\cite{huang2018api, mcmillan2011portfolio, campbell2017nlp2code, allamanis2015bimodal, gvero2015interactive, nguyen2018statistical}. However, these existing works are quite different to the Plot2API since they accomplished the API recommendation tasks based on the source code or textual descriptions. It is not convenient to first convert a plot into textural descriptions or code and then accomplish the task in text to text manner, since the translation of the plot to the code or the textual description leads to the unnecessary time cost and the misinterpretation risk which may target the question to the wrong answer. Instead, the plot-based API recommendation provides an image to text solution which is more intuitional, convenient, and efficient. Nevertheless, to the best of our knowledge, the Plot2API issue remains unstudied. Although the Plot2API issue can be deemed as a common multi-label image classification, it is very challenging due to the extremely varied appearances of the plots drawn by the same API and the unnoticed visualization functions of some subsidiary APIs. Moreover, the APIs also suffer from a serious imbalance which is also a fatal limitation for Plot2API. Figure~\ref{examples} gives some of such examples in the R programming language.

In the recent decade, the Convolutional Neural Networks (CNN) have achieved a significant advance in supervised learning particularly in image classification~\cite{tan2019efficientnet,GoogLeNet,simonyan2014VGG,mlgcn,cnnplp}. They are proficient in learning the discriminative features for images. Here, we leverage a recently advanced CNN model named EfficientNet~\cite{tan2019efficientnet} as the backbone network to develop a novel end-to-end trainable deep learning approach named Semantic Parsing Guided Neural Network  (SPGNN) for filling the aforementioned missing technology. SPGNN introduces an extra semantic parsing module to the EfficientNet which considers the Plot2API issue as a multi-task learning problem for the solution. Besides the conventional EfficientNet-based plot classification flow path, SPGNN extra employs a semantic translation network to translate the visual features of a plot learned from EfficientNet into the semantic representations of APIs and then uses a relation network to compare these estimated semantics with their ground truth for accomplishing the task from the perspective of semantic parsing. By fully exploiting the semantics of APIs, the semantic parsing module facilitates the EfficientNet to better learn the semantic relevant visual features which are more robust to the appearance variation caused by the different parameter settings of the same API. In order to alleviate the sample distribution imbalance of APIs, the random erasing trick is applied to the plots for generating more training data for each category. We release three Plot2API datasets which are collected from Stack Overflow and are carefully preprocessed for evaluating our work. The experimental results show that SPGNN consistently performs better than EfficientNet with a considerable improvement and defeats all deep learning baselines on all datasets.

The main contributions of our work are summarized as follows:

$\bullet$ A novel software engineering task named Plot2API is introduced, which attempts to recommend the graphic APIs based on the plots. Plot2API has many potential and meaningful applications in software engineering.

$\bullet$ A novel deep learning method named Semantic Parsing Guided Neural Network (SPGNN) for tackling the Plot2API task is proposed. SPGNN translates this task into the multi-label image classification and the semantic parsing tasks for the solution. The semantic parsing is expected to facilitate EfficientNet to extract deep features that are more robust to appearance variation and thereby supports the plot-based API recommendation.

$\bullet$ Three novel Plot2API datasets, namely Python-Plot13, R-Plot32 and R-Plot14, are released for evaluation.

$\bullet$ An empirical comparison of classical CNN models on Plot2API is conducted and extensive experimental results on the released datasets demonstrate the superiority of our method over the recent deep learning baselines and its significant improvement over EfficientNet.


\section{Approach}

In this section, we first introduce the Plot2API issue and then elaborate on our proposed method named Semantic Parsing Guided Neural Network (SPGNN).

\subsection{Overview}
\textbf{Problem Formulation:} In this paper, we formulate a new problem in software engineering named Plot2API which studies how to recommend the graphic APIs from plots or figures. According to the facts that each plot may be drawn by multiple APIs and the set of graphic APIs regarding to a programming language is fixed, Plot2API can be deemed as a multi-label image classification task. Let $X = \left\{ x_i|i=1,2,\dots,n \right\}\in \mathcal{R}^{n\times d}$ be the collection of figures and $Y = \left\{ y_i|i=1,2,\dots,n \right\}\in \mathcal{R}^{n\times c}$ be the corresponding labels where $x_i$ is the $i$-th plot and its label $y_i$ is a binary vector. $n$, $d$, and $c$ are the number of samples, the dimension, and the number of APIs respectively. The Plot2API technique aims at learning a mapping function $F(\cdot)$ to map the plots to the labels, i.e.,
\vspace{-0.2cm}
\begin{equation}
  X\overset{F(\cdot)}\rightarrow Y,
\vspace{-0.2cm}
\end{equation}
where $y_i=F(x_i)$. In multi-label image classification, such mapping function is often further divided into two steps, $F(\cdot):=P_\omega( E_\phi(\cdot))$ where $E(\cdot)$ and $P(\cdot)$ are the feature learning and API recommendation respectively. $\phi$ and $\omega$ are their learnable parameters.

\begin{figure*}[t]
\centering
\includegraphics[width=0.88\textwidth,keepaspectratio]{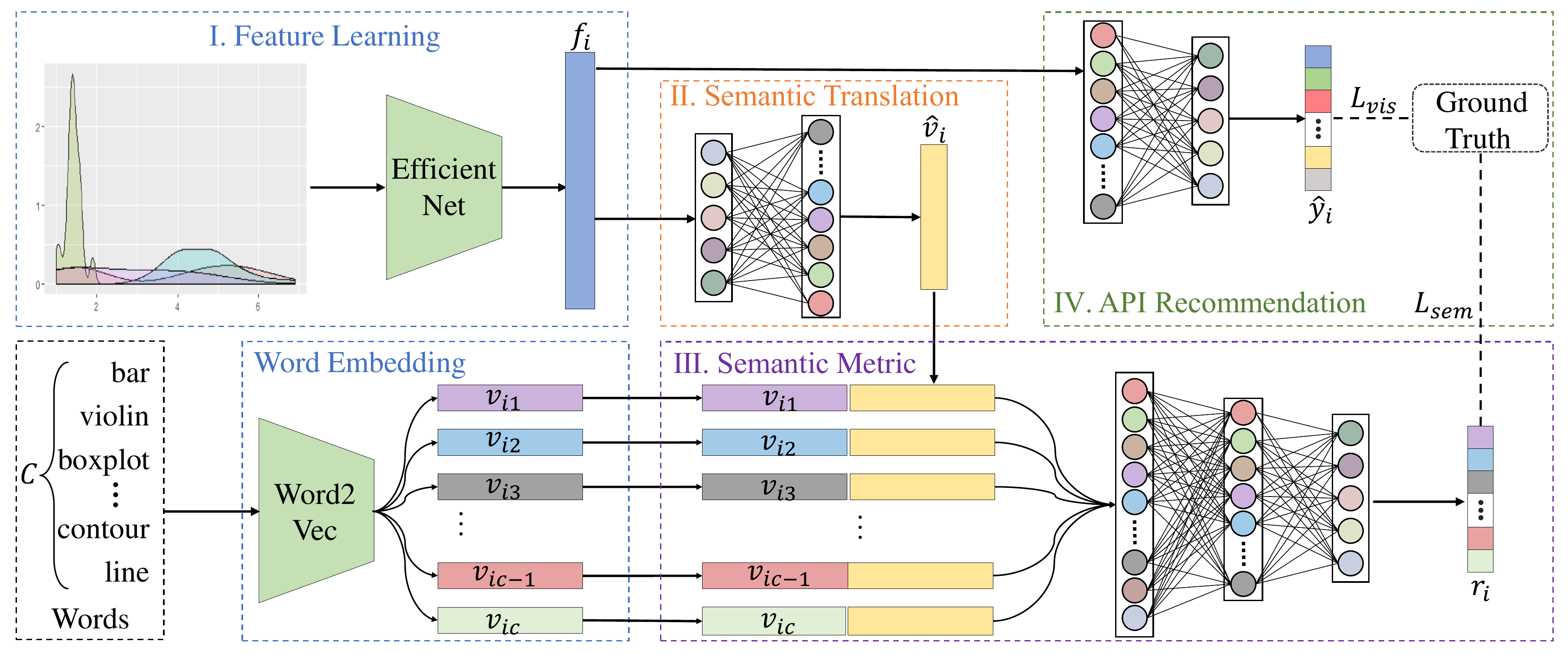}
\vspace{-0.3cm}
\caption{The overview of our method. The input data graphics $x_i$ are sent to the feature learning network to extract the visual features $f_i$, and then generating the predicted API labels $\hat{y_i}$ in API recommendation network and generating the semantic information $\hat{v_i}$ in semantic translation network. The real semantic information is produced by word2vec. After concatenating the semantic vectors, these features are sent to the semantic metric network to evaluate the relational reasoning. The relation is stronger, the output of relation network $r_i$ is more approximate to 1. And then, the $\hat{y_i}$ and $r_i$ are used to recommend APIs in API recommendation network.}
\label{Model}
\vspace{-0.65cm}
\end{figure*}

We consider Plot2API as a multi-label image classification issue for the solution. The recently advanced CNN model is adopted as the backbone of the framework. However, Plot2API is quite different from the original object-based image classification where the samples from the same category often share similar visual appearances. The appearances of the figures drawn by the same API often suffer from the extreme variation since different parameter settings can seriously perplex the plot-based API recommendation, as shown in Figure~\ref{examples}. To overcome this challenge, we intend to utilize the semantics of APIs to guide the feature learning and preserve the semantic relevant visual information which reflects the semantic nature of appearances. Instead of considering the issue as a single-task learning problem, we present a novel deep learning method named Semantic Parsing Guided Neural Network (SPGNN) and regard this issue as a multi-task learning problem for the solution. The merit of this fashion is that relevant tasks can benefit from the solution of each other due to the information complementary. SPGNN contains two relevant tasks namely plot-based API recommendation and plot-based semantic parsing. The plot-based API recommendation is the main task while the plot-based semantic parsing is extra introduced for extracting the semantics of APIs from plots. More specifically, SPGNN consists of feature learning, API recommendation, semantic translation and semantic metric modules. The feature learning and API recommendation modules compose the flow path of plot-based API recommendation while the feature learning, semantic translation and semantic metric modules compose the flow path of plot-based semantic parsing, as shown in Figure~\ref{Model}. The following subsections give the details of these modules.

\subsection{Feature Learning}

In the recent decade, CNN is deemed as the most influential machine learning technique for visual feature learning. Here, we also adopt CNN as the feature learning module. Here, we choose a very recent CNN model named EfficientNet-B3~\cite{tan2019efficientnet} as the feature learning network by considering the trade off between the performance and the efficiency. The empirical study in Section~\ref{exp} also indicates that it is the best performed CNN model for Plot2API. We use the feature extraction network as the mapping function of our feature learning module, which can be denoted as follows,
\vspace{-0.17cm}
\begin{equation}
  f_i = E_\phi(x_i),
  \vspace{-0.17cm}
\end{equation}
where the visual feature $f_i$ is the pooling result of the last convolutional layer's output.

\subsection{Semantic Translation}
The feature learning is the key to the success of the supervised learning model. The single-task learning schema is easy to fall into the overfitting due to the single view of optimization. Here we integrate the semantic parsing module with the aforementioned CNN-based API recommendation flow path and convert such a single task learning issue into a two-task learning issue. The semantic parsing module utilizes a semantic translation network, which consists of one fully connected layer followed by a ReLU layer, to translate the visual feature learned by CNNs into the semantic representation of APIs. Here, we employ the Wikipedia dataset retrained word2vec~\cite{rasiwasia2010new,mikolov2013distributed} to attain the ground truth semantic representation of each API, $V_i=[v_{i1},\cdots,v_{it},\cdots, v_{ic}]$ where $v_{it}$ is a 400-dimensional word embedding corresponding to the $t$-th API and regarding to the $i$-th sample. The semantic translation can be denoted as follows,
\vspace{-0.17cm}
\begin{equation}
  \hat{V}_i = T_\psi(f_i),
  \vspace{-0.17cm}
\end{equation}
where $T(\cdot)$ is the mapping function of the semantic translation network with parameters $\psi$, and $\hat{V}_i=[\hat{v}_{i1},\cdots, \hat{v}_{ic}]$ is the translated semantics of all APIs corresponding to sample $x_i$.

\subsection{Semantic Metric}

By applying the idea of learning to compare~\cite{RN}, we establish a relation network for judging if the translated semantics are identical to the ground truth,
\vspace{-0.2cm}
\begin{equation}
  s_i = R_\vartheta(V_i,\hat{V}_i),
  \vspace{-0.2cm}
\end{equation}
where $s_i$ is a $c$-dimensional semantic relation vector whose $j$-th element $s_i^j$ encodes the semantic relation score between $V_i$ and $\hat{V}_i$. $R(\cdot)$ is the mapping function of the relation network with parameters $\vartheta$ which consists of two fully connected layers followed by ReLU layers. For supervising the semantic translation network to extract the true semantics of APIs, the semantic relation scores should be higher if the corresponding APIs exist in the given figure, and vice versa. We employ the sigmoid function $\sigma(\cdot)$ to normalized the semantic relation scores, $r_i =\sigma (s_i)$, and consider the normalized ones as the occurrence probabilities of APIs in semantics. Then, the above-mentioned target can be reached by measuring the distribution difference between the normalized semantic relation scores and the labels based on the cross-entropy again,
\vspace{-0.2cm}
\begin{equation}
  \mathcal{L}_{sem}= -\sum_{i=1}^N \sum_{j = 1}^{c} y_i^j \log (r_i^j) + (1- y_i^j) \log (1-r_i^j). \label{sem}
  \vspace{-0.2cm}
\end{equation}
By optimizing this loss, the normalized semantic relation score is expected to be 1 or 0 when the figure is not drawn by the corresponding API. Finally, if we rank APIs according to the relation scores, we can obtain a list of API semantics of a plot and then accomplish the plot-based semantic parsing task.

\subsection{API Recommendation}

A one-layer fully connected neural network is leveraged to map the extracted feature $f_i$ into a $c$-dimensional binary label vector. The API recommendation module is denoted as follows,
\vspace{-0.2cm}
\begin{equation}
 \hat{y_i} = P_\omega(f_i),
\vspace{-0.2cm}
\end{equation}
where $P(\cdot)$ is the mapping of the neural network with parameters $\omega$, and $\hat{y_i}$ is a $c$-dimensional predicted label vector whose elements are essentially the estimated occurrence probabilities of the corresponding APIs. In the API recommendation task, we expect to keep the predicted labels be consistent with the ground truth, therefore we adopt the cross-entropy function for measuring such label consistency and denote the label recommendation loss as follows,
\vspace{-0.2cm}
\begin{equation}
\mathcal{L}_{vis} = -\sum_{i=1}^N \sum_{j = 1}^{c} y_i^j \log (\hat{y}_i^j) + (1- y_i^j) \log (1-\hat{y}_i^j), \label{vis}
\vspace{-0.2cm}
\end{equation}
where $y_i^j$ and $\hat{y}_i^j \in [0,1]$ are the label and the predicted occurrence probability of the $j$-th API for the $i$-th sample, and $N$ is the number of samples. Conventionally, for each sample, the graphic APIs are sorted according to $\hat{y}_i$ and then output as the recommendation.

We formulate the model of SPGNN which tackles both the plot-based API recommendation and the plot-based semantic parsing tasks via integrating their losses in Equation~\ref{sem} and \ref{vis},
\begin{equation}
\hat{F} \leftarrow \arg \underset{\phi,\psi,\vartheta,\omega} \min \mathcal{L} := \mathcal{L}_{vis} + \alpha \times \mathcal{L}_{sem},
\end{equation}
where $\hat{F}$ is the trained model and $\alpha$ is a manually tunable positive hyper-parameter for reconciling the losses. After adjusting $\alpha$, we can obtain the trained model.

\subsection{Data Augmentation and API Recommendation}

However, there is a problem that we cannot overlook, that the Plot2API data are extremely imbalanced due to the usage frequency of different APIs. Such imbalance can easily corrupt the supervised learning model. Data augmentation is one of the commonest means for alleviating such problem and also a practical way for avoiding the overfitting. Here, we adopt the random horizontal flips and the very recently proposed data augmentation approach named random erasing \cite{zhong2020random} for enriching the training data of each API. Please note that the semantic parsing module is deemed as a booster, and after the SPGNN model is trained, we only preserve the plot-based API recommendation ﬂow path for API recommendations. Specifically, in testing phase, a plot or figure $x_t$ is input into the feature learning module and then the extracted feature is fed into the API recommendation module for getting its estimated API occurrence probabilities,
\vspace{-0.2cm}
\begin{equation}
  \hat{y}_t = P_\omega(E_\phi(x_t)).
\vspace{-0.2cm}
\end{equation}
Finally, the recommended graphic APIs which are corresponding to the top $k$-highest occurrence probabilities are recommended to this plot.
%

\section{Experimental Setup} \label{exp}
In this section, we first present the three datasets newly released by us. Then, we introduce the evaluation metrics, the implementation details and baselines.


\subsection{Datasets}
To construct datasets for this problem, we first downloaded the Stack Overflow Data Dump of March 2018. Next, we extracted the Python-related and R-related threads from the data dump according to the tags of each thread. Each thread contains a question post and zero or more answer posts. We choose Python and R because they are two popular programming languages and are frequently used for plotting. We further processed the extracted threads, and only kept their posts which are answer posts and contain both image URLs and code. Then, we crawled the images in each answer post from thousands of websites. What's more, the crawled images and extracted code in each post are associated with each other and manually verified by us. The image-code pairs of which the image and code are not matched, the image is not a visualization plot and the code is not Python or R code were removed by us. Finally, we classified the dataset relying on the APIs used in code. To avoid missing and incorrect labels, the labels of each image are manually checked and adjusted by us too.

\subsubsection{Python-Plot13 Dataset\protect\footnotemark[1]}
\begin{figure}[t]
\centering
\includegraphics[width=0.85\columnwidth,keepaspectratio]{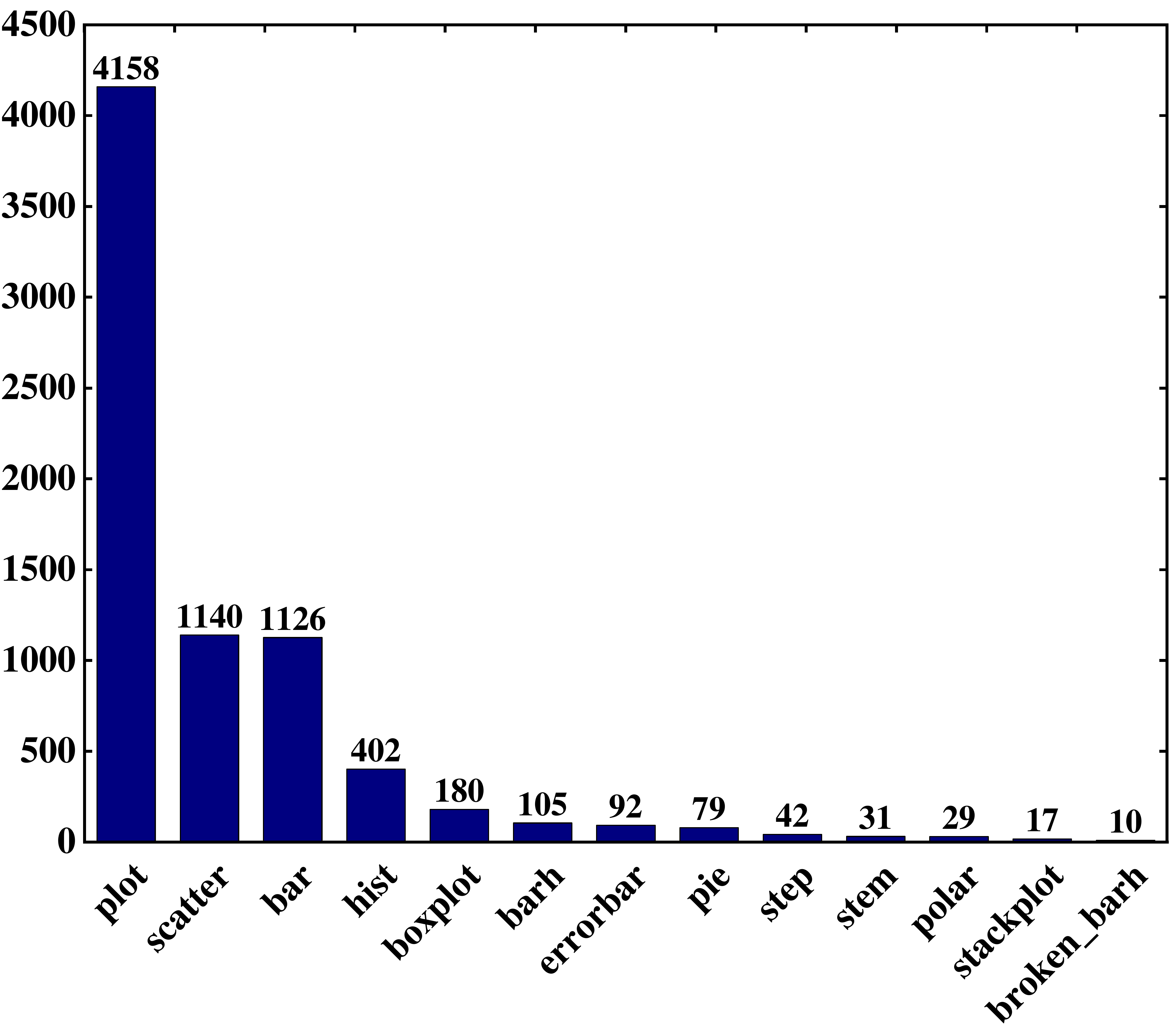} 
\vspace{-0.35cm}
\caption{The data distribution of the Python-Plot13 dataset.}\smallskip
\label{Python}
\vspace{-0.5cm}
\end{figure}

We present a novel Python-based Plot2API dataset named Python-Plot13 dataset. It consists of 6350 python-related plot instances in total and involves 13 APIs, namely bar, barh, boxplot, broken\_barh, errorbar, hist, pie, plot, polar, scatter, stackplot, stem and step. We utilize 5080 samples for training and the rest of 1270 for testing. The data distribution of the Python-Plot13 dataset is shown in Figure~\ref{Python}. From the figure, it is not hard to find that the data are extremely imbalanced. For example, the API plot() possesses more than 4000 instances while broken\_barh() only has 10 instances. Clearly, such imbalance makes the Plot2API very challenging.

\begin{table}[t]
  \centering
  \caption{The data distribution of the R-Plot32 dataset.}
  \vspace{-0.1cm}
  \resizebox{\columnwidth}{!}{
  \begin{tabular}{cc|cc|cc|cc}
    \hline
    API & \# & API & \# & API & \# & API & \# \\
    \hline
    bar & 2111 & bin2d & 12 & density & 205 & density\_2d & 4 \\
    map & 90 & jitter & 105 & boxplot & 638 & quantile & 4 \\
    rug & 20 & smooth & 395 & segment & 385 & contour & 21 \\
    hex & 20 & curve & 8 & dotplot & 40 & errorbar & 335 \\
    step & 38 & line & 2312 & freqpoly & 10 & errorbarh & 39 \\
    sf & 16 &  spoke & 4 & crossbar & 14 & linerange & 49 \\
    path & 232 & violin & 46 & polygon & 303 & pointrange & 49 \\
    point & 3665 & raster & 73 & ribbon & 223 & histogram & 387 \\
    \hline
  \end{tabular}
  }
  \label{R}
  \vspace{-0.6cm}
\end{table}

\subsubsection{R-Plot32 Dataset\protect\footnotemark[1]}The R programming language is regarded as an influential statistical computing language that owns fruitful graphic APIs. Hence, we also propose a new R-based Plot2API dataset named R-Plot32 dataset. The R-Plot32 dataset contains 9114 images where 7292 for training and 1822 for testing. The R-Plot32 involves 32 graphic APIs, namely bar, bin2d, boxplot, contour, crossbar, curve, density, density\_2d, dotplot, errorbar, errorbarh, freqpoly, hex, histogram, jitter, line, linerange, map, path, point, pointrange, polygon, quantile, raster, ribbon, rug, segment, sf, smooth, spoke, step and violin. The number of samples for each API is tabulated in Table~\ref{R}. Similar to the Python-Plot13 dataset, this dataset also suffers from the extreme imbalance of data. Moreover, it is larger and possesses more categories which makes it more challenging than the Python-Plot13 dataset.

\subsubsection{R-Plot14 Dataset\protect\footnotemark[1]}As we can see in Table~\ref{R}, some API functions are used by few images. For example, density\_2d(), spoke() and quantile() classes only have four images in R-Plot32. These functions are used to draw 2D density, directional data points and percentile ratio of total respectively. Besides them, there are some APIs which are rarely used, such as rug(), step(), sf(), curve(), dotplot(), freqpoly(), bin2d(), crossbar() and so on. Hence, we removed these classes from R-Plot32 dataset and construct a reduced version of R-Plot32 named R-Plot14. In addition, there are some APIs belonging to the same super class, such as bar(), errorbar(), errorbarh() and segment(), point(), jitter() and pointrange(), line(), path() and linerange(), which are also removed in R-Plot14. As a supplement dataset, R-Plot14 remove 502 samples(about 5.51\%) from the original 9114 samples, and contains 8612 graphics where 6890 for training and 1722 for testing. The R-Plot14 dataset involves 14 graphic APIs, namely bar, boxplot, contour, density, hex, histogram, line, map, point, polygon, raster, ribbon, smooth, and violin, which have the same images with R-Plot32.

\footnotetext[1]{The datasets and the source code are publicly available at \\ https://github.com/cqu-isse/Plot2API.}




\subsubsection{Data Split Protocol} We randomly select around 80\% of the data to produce the training set while the rest is used as the testing set. In the data split, we ensure that the testing set at least contains one instance for each API.

\subsection{Evaluation Metrics}

We employ Average Precision (AP) as the performance metric for evaluating the recommendation performance for each API. And the AP is essentially the area under the Precision-Recall (P-R) curve which is a popular metric for evaluating the binary classification performances. The mean Average Precision (mAP), known as the mean of APs over all classes, is adopted as a comprehensive metric for evaluating the API-recommendation performance of different methods. The mAP is also known as the commonest metric for multi-label image classification.




%

\subsection{Implementation Details}
We here choose the EfficientNet-B3 \cite{tan2019efficientnet} as our backbone for the trade off between performance and efficiency. Like other deep learning baselines, our backbone network is also pre-trained on ImageNet \cite{deng2009imagenet}. The feature learning module is built from successive MBConv~\cite{tan2019mnasnet, sandler2018mobilenetv2} and convolution layers. After these layers, there is a global average pooling layer. Before being fed into the network, the data graphics will be resized to 300 $\times$ 300. And the dimension of the learned visual feature is 1536. Please refer to the original paper~\cite{tan2019efficientnet} for the detailed architecture of EfficientNet-B3. We adopt word2vec \cite{mikolov2013distributed} trained on the Wikipedia dataset \cite{rasiwasia2010new} to generate the 400-dimensional semantic representations of APIs (word embeddings) for all datasets. Note, the word2vec is retrained, since there is a word (``histogram") not included in the Wikipedia dataset. With regard to the case that an API contains multiple words, we average the embeddings of the words as the API's semantic representation. The semantic translation network and API recommendation network all consist of just one fully connected layer while the relation network is a neural network with two fully connected layers, whose hidden layer is 256.



We train the proposed model using an Adam optimizer \cite{kingma2014adam} with the batch size of 32 and momentum of 0.99. ReLU is used as the activation function in all the fully connected layers. The network is trained for 100 epochs in total. We implement the network based on PyTorch.

\subsection{Baselines}
VGG-16~\cite{simonyan2014VGG}, ResNet-50~\cite{he2016Resnet}, Inception-V1~\cite{GoogLeNet}, and EfficientNet-B3~\cite{tan2019efficientnet} are deemed as representative deep learning approaches for image classification and are regarded as the baseline methods. The main contribution of VGGNet is the increased depth with very small convolution filters~\cite{simonyan2014VGG}. ResNet utilized a residual network, which is easy to optimize, to improve the accuracy from considerably increased depth~\cite{he2016Resnet}. To improve the utilization of the computing resources, Inception was proposed as a sparse structure by readily available dense building blocks to improve neural networks for computer vision~\cite{GoogLeNet}. EfficientNet balanced network depth, width, and resolution to lead a better performance than other CNNs~\cite{tan2019efficientnet}.

\section{Experimental Results}
In this section, we conduct experiments to evaluate our proposed model on three datasets. Then, we carry out ablation studies to evaluate the effectiveness of the proposed module in SPGNN. The goal of experimental results shown in this section is to answer the following questions:

$\bullet$ \textbf{RQ1}: How effective is SPGNN for API recommendation?

$\bullet$ \textbf{RQ2}: How well do our SPGNN model perform after combining the semantic parsing module and the random erasing-based data augmentation?

$\bullet$ \textbf{RQ3}: How well do our SPGNN model perform when training and testing across different programming languages?

\subsection{RQ1: How eﬀective is SPGNN for API recommendation?}

\begin{table}[t]
  \centering
  \caption{The performance comparison on all datasets. }
  \vspace{-0.1cm}
  \normalsize{
  \resizebox{\columnwidth}{!}{
  \begin{tabular}{cccc}
    \hline
    \diagbox{mAP}{Datasets} & Python-Plot13 & R-Plot32 & R-Plot14 \\
    \hline
    VGG-16 & 67.46 & 38.39 & 66.08 \\
    VGG-16 + DA & 64.10 & 40.84 & 67.96 \\
    ResNet-50 & 56.33 & 29.64 & 55.81 \\
    ResNet-50 + DA & 55.95 & 29.81 & 56.53 \\
    Inception-v1 & 52.92 & 26.06 & 51.84 \\
    Inception-v1  + DA & 54.93 & 32.59 & 53.41 \\
    EfficientNet-B3 & 68.51 & 44.61 & 70.75 \\
    EfficientNet-B3 + DA & 69.33 & 44.46 & 71.29 \\
    \hline
    \textbf{SPGNN} & 71.16 & 45.63 & 71.84 \\
    \textbf{SPGNN + DA} & \textbf{75.95} & \textbf{47.76} & \textbf{75.13} \\
    \hline
  \end{tabular}
  }}
 \vspace{-0.6cm}
  \label{mAP}
\end{table}
%

We compare SPGNN with four well-known image classification approaches, including VGG-16~\cite{simonyan2014VGG}, ResNet-50~\cite{he2016Resnet}, Inception-V1~\cite{GoogLeNet}, and EfficientNet-B3~\cite{tan2019efficientnet} on our datasets. Table~\ref{mAP} tabulates the mAP of different methods on different datasets. Tables~\ref{pythonAP}, \ref{RAP} and \ref{subRAP} report the AP of each API on Python-Plot13, R-Plot32 and R-Plot14 datasets respectively. Clearly, EfficientNet-B3 significantly outperforms VGG-16, ResNet-50 and Inception-V1 on all datasets. Therefore, we choose the EfficientNet-B3 as our backbone. From observations, it is not hard to find that our proposed model consistently performs much better than state-of-the-art CNN approaches and achieves considerable mAP improvement over EfficientNet-B3 which is our baseline on all datasets. Here, we will present the detail experimental analysis individually.

\begin{table*}[t]
  \centering
  \caption{The performance comparison on the Python-Plot13 dataset (the AP for each category while the mAP for all, the bold number indicates the best performance and DA = random erasing-based data augmentation ).}
  \vspace{-0.2cm}
 \resizebox{\textwidth}{!}{
 \begin{tabular}{ccccccccccccccc}
    \hline
    Methods & \textbf{mAP} & bar & barh & boxplot & broken\_barh & errorbar & hist & pie & plot & polar & scatter & stackplot & stem & step \\
    \hline
    VGG-16 & 67.46 & 85.02 & 45.33 & 95.11 & 50.29 & 55.29 & 71.62 & 91.16 & 92.97 & 66.11 & \textbf{80.39} & 55.80 & \textbf{66.90} & 20.98 \\
    VGG-16 + DA & 64.10 & 85.72 & 46.57 & 96.10 & 6.87 & 56.44 & 71.45 & 91.69 & 93.61 & 73.65 & 77.66 & 66.79 & 25.33 & 41.50 \\
    ResNet-50 & 56.33 & 79.04 & 44.33 & 84.56 & 0.58 & 24.13 & 55.63 & 98.66 & 92.66 & 70.41 & 74.76 & 38.36 & 34.56 & 34.66 \\
    ResNet-50 + DA & 55.95 & 80.29 & 47.59 & 89.49 & 1.30 & 24.72 & 54.17 & 99.36 & 92.89 & 70.33 & 76.01 & 40.32 & 35.22 & 15.63 \\
    Inception-V1 & 52.92 & 82.91 & 42.77 & 90.41 & 4.61 & 17.59 & 56.17 & \textbf{100.00} & 90.47 & 55.68 & 74.79 & 50.11 & 6.56 & 15.91 \\
    Inception-V1 + DA & 54.93 & 83.64 & 42.66 & 88.49 & 1.41 & 17.46 & 62.87 & 96.10 & 91.67 & 67.42 & 75.01 & 47.98 & 21.06 & 18.34 \\
    EfficientNet-B3 & 68.51 & 87.67 & 53.36 & 97.82 & 1.72 & 58.66 & \textbf{78.47} & \textbf{100.00} & 93.53 & 68.00 & 77.82 & 74.36 & 66.75 & 32.48 \\
    EfficientNet-B3 + DA & 69.33 & \textbf{88.53} & 49.02 & \textbf{97.91} & 75.00 & 55.88 & 65.52 & \textbf{100.00} & 92.87 & \textbf{82.23} & 79.13 & 47.39 & 33.63 & 34.19 \\
    \hline
    \textbf{SPGNN} & 71.16 & 86.57 & 54.68 & 95.85 & 4.32 & \textbf{71.98} & 73.50 & \textbf{100.00} & 94.00 & 79.29 & 79.12 & 75.76 & 66.85 & \textbf{43.15} \\

    \textbf{SPGNN + DA} & \textbf{75.95} & 86.15 & \textbf{56.76} & 96.72 & \textbf{100.00} & 55.71 & 77.50 & 93.41 & \textbf{94.08} & 62.93 & 80.37 & \textbf{80.95} & 66.81 & 35.97 \\
    \hline
  \end{tabular}
  }
 \vspace{-0.4cm}
  \label{pythonAP}
\end{table*}

\begin{table*}[t]
  \centering
  \caption{The performance comparison on the R-Plot32 dataset (the AP for each category while the mAP for all, the bold number indicates the best performance and DA = random erasing-based data augmentation).}
  \vspace{-0.2cm}
  \resizebox{\textwidth}{!}{
  \begin{tabular}{cccccccccccccccccc}
    \hline
    Methods & \textbf{mAP} & bar & bin2d & boxplot & contour & crossbar & curve & density & density\_2d & dotplot & errorbar & errorbarh & freqpoly & hex & histogram & jitter & line  \\
    \hline
    VGG-16 & 38.39 & 92.09 & 7.22 & 92.14 & \textbf{31.48} & 0.69 & 33.78 & 79.05 & \textbf{100.00} & 17.89 & 65.73 & 46.44 & 0.72 & 42.20 & 50.88 & 10.24 & 83.80 \\
    VGG-16 + DA & 40.84 & 93.96 & 8.22 & 91.43 & 12.92 & 3.42 & 0.46 & 82.17 & 0.28 & 30.59 & 75.26 & \textbf{56.52} & 0.32 & 43.49 & 63.33 & 15.25 & 87.31 \\
    ResNet-50 & 29.64 & 92.31 & 0.80 & 83.25 & 14.94 & 5.79 & 5.12 & 69.49 & 0.44 & 11.57 & 39.05 & 41.09 & 0.79 & 41.16 & 41.35 & 13.63 & 79.73 \\
    ResNet-50 + DA & 29.81 & 92.05 & 1.71 & 83.37 & 12.84 & 4.01 & 4.46 & 71.83 & 0.69 & 12.75 & 40.66 & 53.58 & 0.66 & 26.10 & 44.13 & 12.72 & 79.91 \\
    Inception-V1 & 26.06 & 90.80 & 1.24 & 86.61 & 0.79 & 0.42 & 0.84 & 56.00 & 2.63 & 1.58 & 35.48 & 15.79 & 0.41 & 67.33 & 46.21 & 6.01 & 80.38 \\
    Inception-V1 + DA & 32.59 & 92.96 & 4.26 & 89.36 & 1.91 & \textbf{17.04} & 0.33 & 69.66 & \textbf{100.00} & 1.73 & 51.69 & 26.44 & 0.16 & 34.40 & 50.89 & 7.58 &  81.21 \\
    EfficientNet-B3 & 44.61 & 94.82 & 29.08 & 92.55 & 6.39 & 1.85 & 13.61 & 87.66 & \textbf{100.00} & 50.07 & 71.56 & 49.74 & 0.72 & 47.89 & \textbf{69.99} & 19.68 & 87.54 \\
    EfficientNet-B3 + DA & 44.46 & 95.38 & \textbf{31.94} & 92.86 & 12.40 & 0.36 & 17.23 & \textbf{92.00} & 3.33 & 32.44 & 71.14 & 54.44 & \textbf{5.15} & 35.37 & 67.10 & \textbf{34.03} & 87.02 \\
    \hline
    \textbf{SPGNN} & 45.63 & 92.55 & 2.00 & \textbf{94.10} & 10.67 & 2.63 & \textbf{34.19} & 85.69 & \textbf{100.00} & \textbf{51.66} & \textbf{77.56} & 47.02 & 2.74 & 64.65 & 63.32 & 13.26 & \textbf{87.95} \\
    \textbf{SPGNN + DA} & \textbf{47.76} & \textbf{95.96} & 4.01 & 91.74 & 8.83 & 1.32 & \textbf{34.49} & 88.69 & \textbf{100.00} & 47.61 & 75.36 & 41.49 & 0.31 & \textbf{68.55} & 69.57 & 26.71 & 86.43 \\
    \hline
    \hline
    Methods & map & path & point & pointrange & polygon & quantile & raster & ribbon & rug & segment & sf & smooth & spoke & step & violin & linerange & -  \\
    \hline
    VGG-16 & 34.55 & 12.96 & 94.68 & 31.05 & 44.09 & 0.35 & 41.97 & 40.16 & 18.09 & 15.88 & 10.07 & 42.14 & 0.31 & 34.42 & 26.97 & 26.45 & -  \\
    VGG-16 + DA & 37.60 & 20.70 & 93.68 & 61.31 & 43.70 & 0.20 & 45.63 & 42.03 & 19.93 & 20.43 & 10.16 & 49.06 & 0.39 & 35.66 & 31.24 & 16.94 & -  \\
    ResNet-50 & 48.33 & 18.81 & 87.94 & 19.59 & 43.14 & 0.32 & 47.27 & 27.41 & 9.09 & 9.18 & \textbf{13.66} & 39.53 & 0.31 & 28.92 & 11.28 & 3.29 & - \\
    ResNet-50 + DA  & 37.82 & 19.02 & 88.28 & 22.37 & 44.47 & \textbf{0.36} & 42.76 & 28.13 & 8.28 & 9.16 & 8.40 & 39.42 & 0.28 & 22.15 & 11.08 & 2.51 & - \\
    Inception-V1 & 31.77 & 9.05 & 89.83 & 32.07 & 42.32 & 0.25 & 8.52 & 41.08 & 0.50 & 11.46 & 3.83 & 37.11 & 0.11 & 20.61 & 7.98 & 5.00 & - \\
    Inception-V1 + DA & 44.70 & 8.66 & 89.92 & 43.96 & 43.29 & 0.22 & 38.93 & 39.22 & 3.69 & 15.44 & 0.53 & 37.28 & 0.21 & 37.85 & 7.41 & 1.85 & - \\
    EfficientNet-B3 & 49.80 & \textbf{29.90} & 94.54 & 41.42 & 47.68 & 0.08 & \textbf{60.25} & 64.86 & 31.41 & 26.77 & 2.86 & 50.70 & 6.82 & 26.90 & 40.36 & 30.02 & - \\
    EfficientNet-B3 + DA & 48.90 & 26.66 & \textbf{95.46} & 48.76 & 56.90 & 0.34 & 50.93 & 59.81 & 29.15 & 26.81 & 1.17 & 50.76 & \textbf{70.00} & 43.18 & 51.47 & 30.17 & - \\
    \hline
    \textbf{SPGNN} & 40.61 & 22.18 & 94.84 & \textbf{79.04} & 42.72 & 0.17 & 49.02 & 63.89 & 27.03 & \textbf{26.92} & 0.80 & \textbf{52.29} & 0.46 & 31.40 & 57.21 & 41.71 & - \\
    \textbf{SPGNN +DA} & \textbf{60.50} & 23.40 & 95.13 & 64.05 & \textbf{58.65} & 0.20 & 48.88 & \textbf{65.57} & \textbf{44.15} & 23.96 & 7.06 & 50.97 & 0.94 & \textbf{38.11} & \textbf{60.08} & \textbf{45.73} & - \\
    \hline
  \end{tabular}
  }
 \vspace{-0.4cm}
  \label{RAP}
\end{table*}

\subsubsection{Results on Python-Plot13 dataset} SPGNN and SPGNN+DA respectively achieve 71.76\% and 75.95\% mAP and perform the best in comparison with all baselines. The performance gains of SPGNN+DA over VGG-16, ResNet-50, Inception-V1 and EfficientNet-B3 in mAP are 8.49\%, 19.62\%, 23.03\% and 7.44\% respectively. After introducing the same data augmentation to these four baselines, our method still demonstrates the significant advantages over these methods and the performances gains are 11.85\%, 20.00\%, 21.02\% and 6.62\% respectively. Moreover, it also worthwhile to point out DA is not always work for all CNNs. For examples, VGG-16 and ResNet-50 with DA are performs much worse than their original versions on Python-Plot14 dataset.

According to Table~\ref{pythonAP}, our model also gets the first on the API recommendations of barh(), broken\_barh(), errorbar(), pie(), plot(), stackplot() and step() APIs among all 13 APIs. Particularly, our model gets 100\% AP in broken\_barh() prediction where such numbers of VGG-16, ResNet-50, Inception-V1 and EfficientNet-B3 are only 50.29\%, 0.58\%, 4.61\% and 1.72\% respectively. In bar(), boxplot(), scatter() and stem(), the performance of our model is very close to the first one. More than half of APIs get over 80\% AP via our model. This implies that SPGNN possesses the good potential for Python graphic API recommendation in reality. Moreover, the experimental results demonstrate that the random erasing-based data augmentation improves SPGNN by the mAP of 4.79\% and makes SPGNN become more balanced cross all APIs. We attribute these to the fact that the random erasing-based data augmentation enriches the appearances of plots and mitigates the overfitting of the proposed model.

All the methods do not perform well on the API recommendation of barh(), errorbar(), polar(), stem() and step(). The reason behind this phenomenon we believe is that the appearances of the plots drawn by barh() and errorbar() are extremely similar, since barh() and errorbar() are both variants of bar(), while the figures drawn by polar() have the similar appearance with pie(), which both contain the circle element. The graphics plotted by step() share the similar feature with bar() and the plots drawn by stem() have the visual features of point() and line(). What's more, the number of step() and stem() is only 42 and 31 in the dataset, which limits the learning power of CNNs to a certain extent. Even so, by incorporating the semantic information of APIs, SPGNN still significantly improves the performance of the recommendation of these APIs.

\begin{figure}[t]
\centering
 \centering
 \subfigure[Python graphic example]{
   \includegraphics[width=0.98\columnwidth]{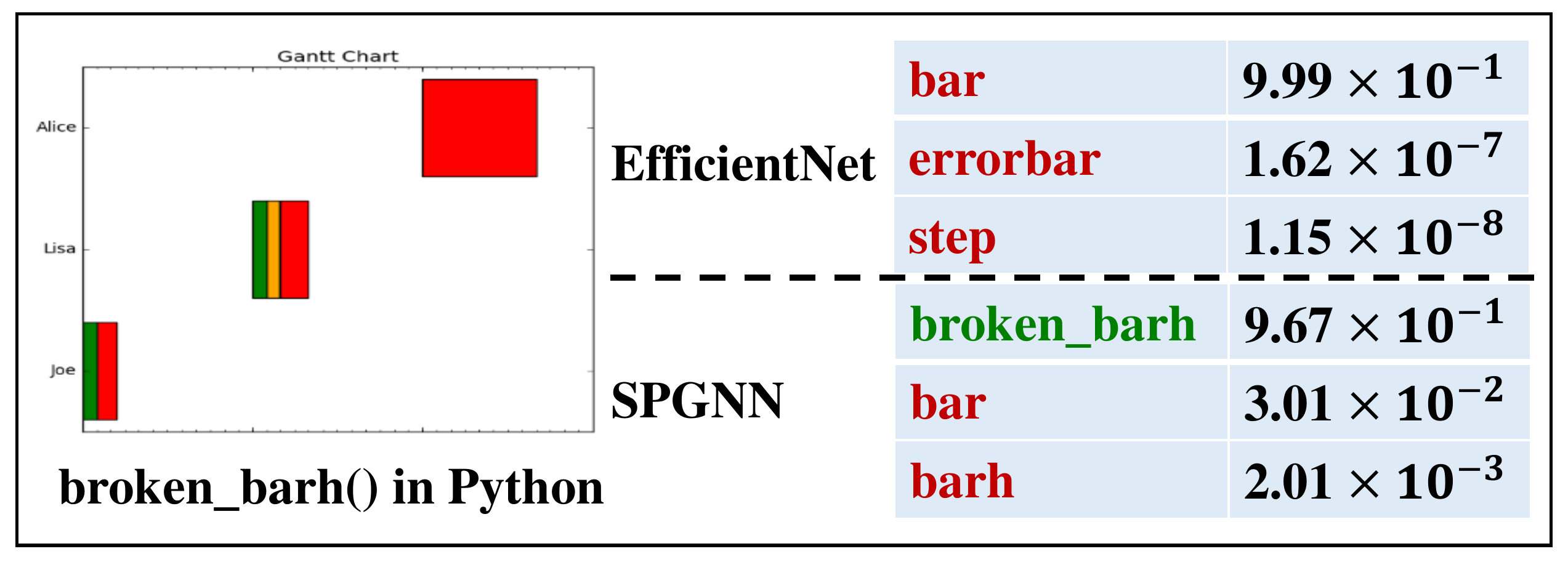}
   \label{python-case}
 }
 \subfigure[R graphic example]{
   \includegraphics[width=0.98\columnwidth]{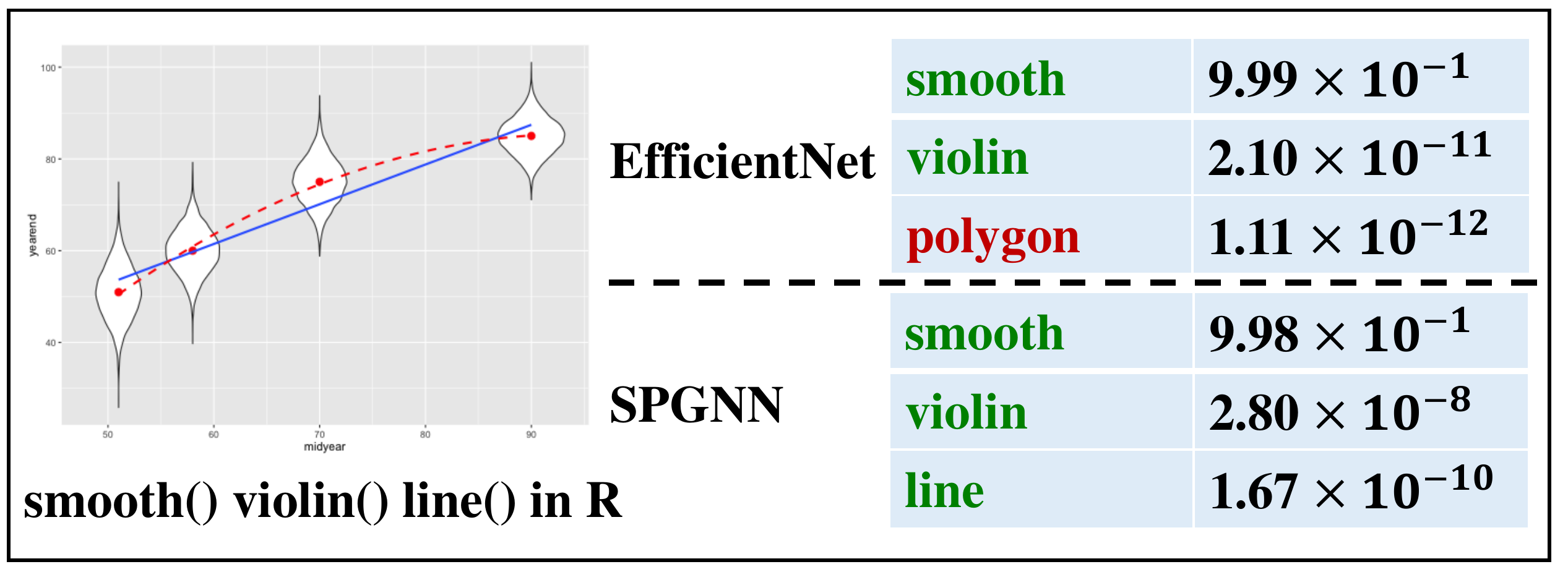}
   \label{R32-case}
 }
\vspace{-0.4cm}
\caption{The Python and R graphic API recommendation examples. The top-3 APIs recommended by EfficientNet and our method via giving the Python or R based plots. The green ones are the correct APIs while the red ones are the wrong API.}\label{cases}
\vspace{-0.75cm}
\end{figure}

\begin{figure}[b]
\vspace{-0.7cm}
\centering
\includegraphics[width=0.7\columnwidth]{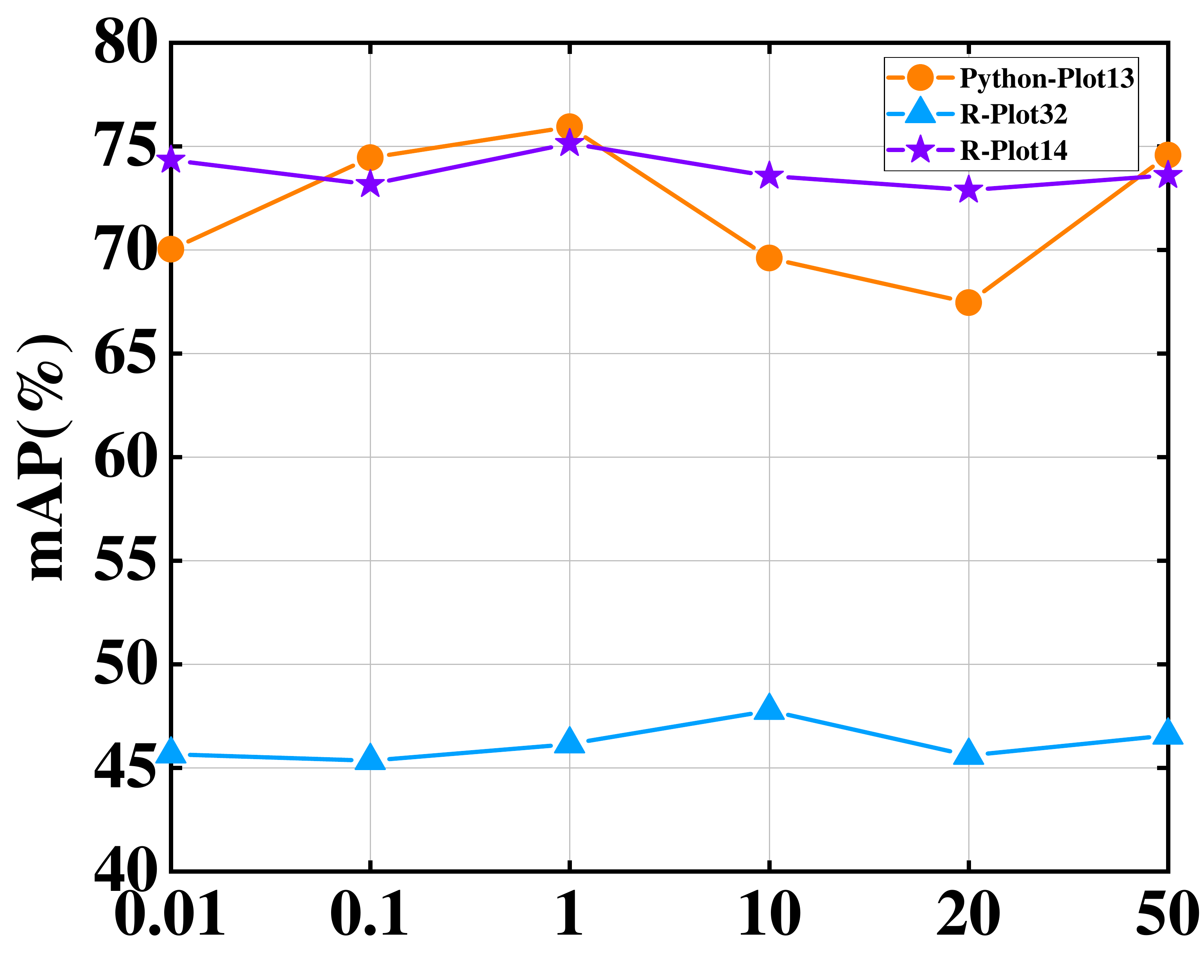}
\vspace{-0.3cm}
\caption{The influence of hyper-parameter $\alpha$ to the performance of SPGNN (in mAP).}
\label{alpha}
\vspace{-0.4cm}
\end{figure}

\begin{table*}[h]
  \centering
  \caption{The performance comparison on the R-Plot14 dataset(the AP for each category while the mAP for all, the bold number indicates the best performance and DA = random erasing-based data augmentation).}
  \vspace{-0.2cm}
  \resizebox{\textwidth}{!}{
  \begin{tabular}{cccccccccccccccc}
    \hline
    Methods & \textbf{mAP} & bar & boxplot & contour & density & hex & histogram & line & map & point & polygon & raster & ribbon & smooth & violin \\
    \hline
    VGG-16 & 66.08 & 95.09 & 93.10 & 33.26 & 81.24 & 59.19 & 59.10 & 88.98 & 43.24 & 95.02 & 52.35 & 60.62 & 36.28 & 52.16 & 75.46 \\
    VGG-16 + DA & 67.96 & 93.93 & 94.42 & 23.82 & 78.75 & 53.28 & 70.69 & 88.03 & 45.15 & 94.61 & 48.39 & 69.73 & 43.02 & 58.57 & 89.08 \\
    ResNet-50 & 55.81 & 91.57 & 88.32 & 20.97 & 71.17 & 38.48 & 51.85 & 83.77 & 47.41 & 90.66 & 45.23 & 34.48 & 25.57 & 50.59 & 41.32 \\
    ResNet-50 + DA & 56.53 & 91.49 & 89.33 & 22.93 & 71.35 & 41.76 & 50.79 & 83.48 & 46.66 & 90.54 & 45.29 & 37.50 & 26.74 & 50.71 & 42.79 \\
    Inception-V1 & 51.84 & 93.54 & 86.20 & 20.01 & 57.59 & 32.64 & 57.72 & 79.96 & 30.34 & 92.24 & 30.88 & 25.96 & 29.25 & 45.09 & 44.29 \\
    Inception-V1 + DA & 53.41 & 93.28 & 90.14 & 32.89 & 54.12 & 18.29 & 56.70 & 83.89 & 32.11 & 93.20 & 39.08 & 28.60 & 35.01 & 51.24 & 39.19 \\
    EfficientNet-B3 & 70.75 & \textbf{96.23} & 97.28 & 23.29 & 84.52 & \textbf{76.86} & 74.93 & 90.17 & \textbf{54.83} & 95.56 & 56.00 & 62.84 & 42.31 & 60.35 & 75.37 \\
    EfficientNet-B3 + DA & 71.29 & 93.81 & \textbf{97.78} & 25.37 & 83.30 & 39.49 & \textbf{81.64} & \textbf{91.85} & 53.99 & \textbf{95.98} & 55.64 & 70.79 & 51.81 & 61.88 & \textbf{94.76} \\
    \hline
    \textbf{SPGNN} & 71.84 & 95.83 & 97.03 & 29.36 & \textbf{87.61} & 63.58 & 79.39 & 90.27 & 52.91 & 95.52 & \textbf{59.66} & 63.08 & 46.68 & 67.56 & 77.24 \\
    \textbf{SPGNN + DA} & \textbf{75.13} & 95.04 & 96.61 & \textbf{39.87} & 84.20 & 75.09 & 80.55 & 90.42 & 51.37 & 95.81 & 54.96 & \textbf{75.79} & \textbf{55.05} & \textbf{69.41} & 87.72 \\
    \hline
  \end{tabular}
  }
  \label{subRAP}
  \vspace{-0.6cm}
\end{table*}

\subsubsection{Results on R-Plot32 dataset} The R-Plot32 dataset is a more challenging dataset with more samples and more APIs. Our method still performs the best. The gains of SPGNN over VGG-16, ResNet-50, Inception-V1 and EfficientNet-B3 in mAP are 7.24\%, 15.99\%, 19.57\% and 1.02\%  respectively and such numbers of SPGNN+DA are 9.37\%, 18.12\%, 21.70\% and 3.15\%. The improvements of SPGNN+DA over baselines+DA are 6.92\%, 17.95\%, 15.17\% and 3.30\%. Moreover, our method also achieves the first rank 18 times among 32 APIs.

According to the results, many similar phenomena on the Python-Plot13 dataset are also observed on the R-Plot32 dataset. Here, we will not give the same conclusions introduced in the previous section and only focus on analyzing the phenomena specific to the R-Plot32 dataset. The most obvious phenomenon is that almost all methods fail on the recommendation of some APIs, such as bin2d(), contour(), crossbar(), freqpoly(), qunatitle(), sf() and spoke(). We believe the reason behind this is the small training size of these APIs limits the learning power of CNN. For example, sf(), spoke() and qunatitle() only have 4 samples in total. Additionally, the figures or shape appearances drawn by the subsidiary APIs, such as contour() and spoke(), often highly relate to the appearances of the main objects in the plot or only cover a tiny fraction of the figure which is hard to be visually noticed. It is also difficult to distinguish the figures drawn by APIs like bin2d() and crossbar(), since some other APIs can draw very similar figures. For example, the figure drawn by bin2d() can be easily identified as a rectangle, and there are many graphic APIs in R, such as bar() and line(), can draw the rectangle-like shapes.

Although our method performs fairly well on some frequently used APIs, such as line(), point() and bar(), there exists a large gap between the Python-Plot13 and the R-Plot32 datasets in terms of the overall performance measured by mAP. The main reason of such low mAP we believe is the lack of sufficient training data for some APIs. Hence, we have also conducted several experiments on a reduced version of the R-Plot32 dataset, namely R-Plot14, for validating the effects of our methods in the case that each API contains enough training data.

\subsubsection{Results on R-Plot14 dataset}
The results on R-Plot14 are shown in Table~\ref{subRAP}. We find in surprise that all methods' performance is significantly boosted on the R-Plot14 dataset, which removed the similar APIs and the classes with few graphics. Since the classes have obvious distinguishing features, our model demonstrates a better performance (+27.37\%) on R-Plot14 compared with R-Plot32, which is very similar to that of the Python-Plot13 dataset. This phenomenon reflects the application possibility of our method on R programming language in the future.


As we can see from Table~\ref{subRAP}, the performance of most APIs is boosted compared with the R-Plot32 dataset. SPGNN+DA gets 75.13\% in mAP, which is 9.05\%, 19.32\%, 23.29\% and 4.38\% higher than VGG-16, ResNet-50, Inception-V1, and EfficientNet-B3, and also gets a better performance than baselines+DA about 7.17\%, 18.60\%, 21.72\% and 3.84\%. Specifically, among 14 APIs, our approach achieves the best API recommendation performance on contour(), density(), polygon(), raster(), ribbon(), and smooth(). As for the other APIs, EfficientNet-B3 achieves the best performance, but our model has the little gap with it.


The performance of contour() is not good among all the methods on the API recommendation because there are only 24 samples in total, but our model still performs best via all the methods. After getting more training data, we believe that the performance will be better. It is also worthwhile to point out that data augmentation trick significantly improves the recommendation performance of contour(). This implies that the random erasing-based data augmentation indeed alleviate the imbalance of sample across the categories, particulary can benefit the recommendation of API which owns limited samples.

\subsubsection{Some Successful Plot2API Examples of SPGNN}
Figure~\ref{cases} shows two cases that our method obtains a better API recommendation over EfficientNet on Python and R plots. In Figure~\ref{python-case}, SPGNN gets the right python API label as the first recommendation with the confidence of 0.97 while EfficientNet fails. Figure~\ref{R32-case} indicates that SPGNN finds all three correct R graphic APIs while EfficientNet misses the line().


\vspace{-0.3cm}\RS{1}{The SPGNN outperforms the state-of-the-art baselines VGG-16, ResNet-50, Inception-v1 and EfficientNet-B3 substantially on the respect of API recommendation. The results reflect that our model is effective and can be used to assist developers in plotting.}\vspace{-0.5cm}

\subsection{RQ2: How well do our SPGNN model perform after combining the semantic parsing module and the random erasing-based data augmentation?}

The EfficientNet-B3 can be deemed as the plain version of SPGNN without the semantic parsing. From the observations in Table~\ref{pythonAP}, \ref{RAP} and \ref{subRAP}, SPGNN are consistently better than EfficientNet-B3 on all three datasets. More specifically, the mAP improvements of SPGNN over EfficientNet-B3 are 2.65\%, 1.02\% and 1.09\% on Python-Plot13, R-Plot32 and R-Plot14 datasets respectively. Moreover, these observations also demonstrate the considerable improvement of the off-the-shelf data augmentation trick on SPGNN. As we can see from Table~\ref{pythonAP}, \ref{RAP} and \ref{subRAP}, the performance of SPGNN+DA are 4.79\%, 2.13\% and 3.29\% higher than SPGNN.

SPGNN only involves one manually tunable parameter $\alpha$, which is used to reconcile the optimization of the involved two tasks. A greater $\alpha$ means to pay more attention on the solution of the semantic parsing task. Figure~\ref{alpha} shows the impacts of different $\alpha$ on the performance of SPGNN. According to the results, the best $\alpha$ is 1, 10 and 1 on Python-Plot13, R-Plot32 and R-Plot14 datasets respectively, which means the visual features and semantic features have the similar weight in our model.

\vspace{-0.3cm}\RS{2}{The semantic parsing and data augmentation modules are two important parts of our model. After composing these two tricks, the performance confirms the effectiveness of these modules for the API recommendation.}\vspace{-0.5cm}

\subsection{RQ3: How well do our SPGNN model perform when training and testing across different programming languages?}
 \vspace{-0.6cm}
 \begin{table}[htbp]
  \centering
  \caption{The cross-language API recommendation performances of SPGNN in mAP. }
  \vspace{-0.3cm}
 \begin{tabular}{c|ccc}
    \hline
    \diagbox{Datasets}{APIs} & bar & boxplot & plot/line \\
    \hline
    Python-Plot13 & 87.07 & 84.85 &	96.70 \\
    R-Plot32 & 93.65 & 82.20 & 93.04 \\
    R-Plot13 & 90.40 & 89.74 & 97.54 \\
    \hline
  \end{tabular}
 \vspace{-0.3cm}
  \label{cross-language}
\end{table}

In order to evaluate the effectiveness of our method in dealing with the cross-language API recommendation, ten developers independently pick up the shared APIs in Python-Plot13, R-Plot32 and R-Plot14, namely bar(), boxplot() and plot(), which is called line() in R programming language. In these experiments, we employ the data of one programming language for training our model while the data of the other programming language is used for testing. Table~\ref{cross-language} records such experimental results. Taking the first row of results as an example, we train our model on Python-Plot13 dataset, and test the model using the figures plotted by R programming language. In such case, the recommendation accurcies of bar(), boxplot() and plot() are 87.07\%, 84.85\% and 96.70\% respectively. With regard to the experiments related to the last two rows of results, the data of R-Plot32 and R-Plot14 are used for trained respectively, while the samples of Python-Plot13 related to the involved APIs are used for testing. The observations on the last two rows of Table~\ref{cross-language} show that our method still obtains the similarly good results. Moreover, the recommendation performance of these three APIs via using our model trained in a cross language way is very similar to the one observed in Tables~\ref{pythonAP}, \ref{RAP} and \ref{subRAP}, which are the results produced by our model trained in normal way. These phenomena all imply that SPGNN essentially learns the structural geometric characteristics of plots across different languages and it is possible to conduct the cross-language API conversion based on the plots.


\vspace{-0.3cm}\RS{3}{Our model shows the effectiveness of cross-language API recommendation. No matter what language is used for plotting, it can recommend the APIs of Python and R programming languages successfully, as long as similar features shared among the graphics.}\vspace{-0.5cm}

\section{Discussion}

In this section, we first present the real-world user scenarios. Then, threats to validity will be introduced.

\subsection{Usage Scenarios}

\begin{figure}[t]
 \centering
 \subfigure[The Plot2API example of web figure]{
   \includegraphics[width=0.9\columnwidth]{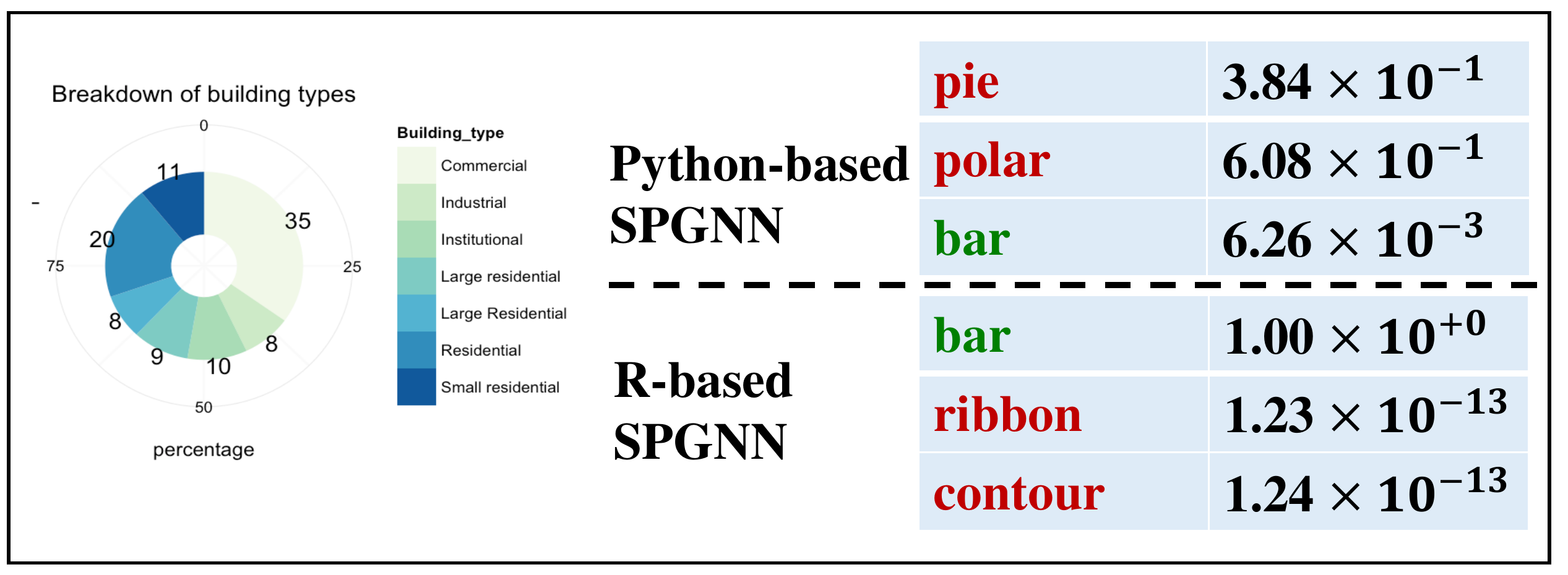}
   \label{eg1}
 }
 \centering
 \subfigure[The Plot2API example of hand-drawn figure]{
   \includegraphics[width=0.9\columnwidth]{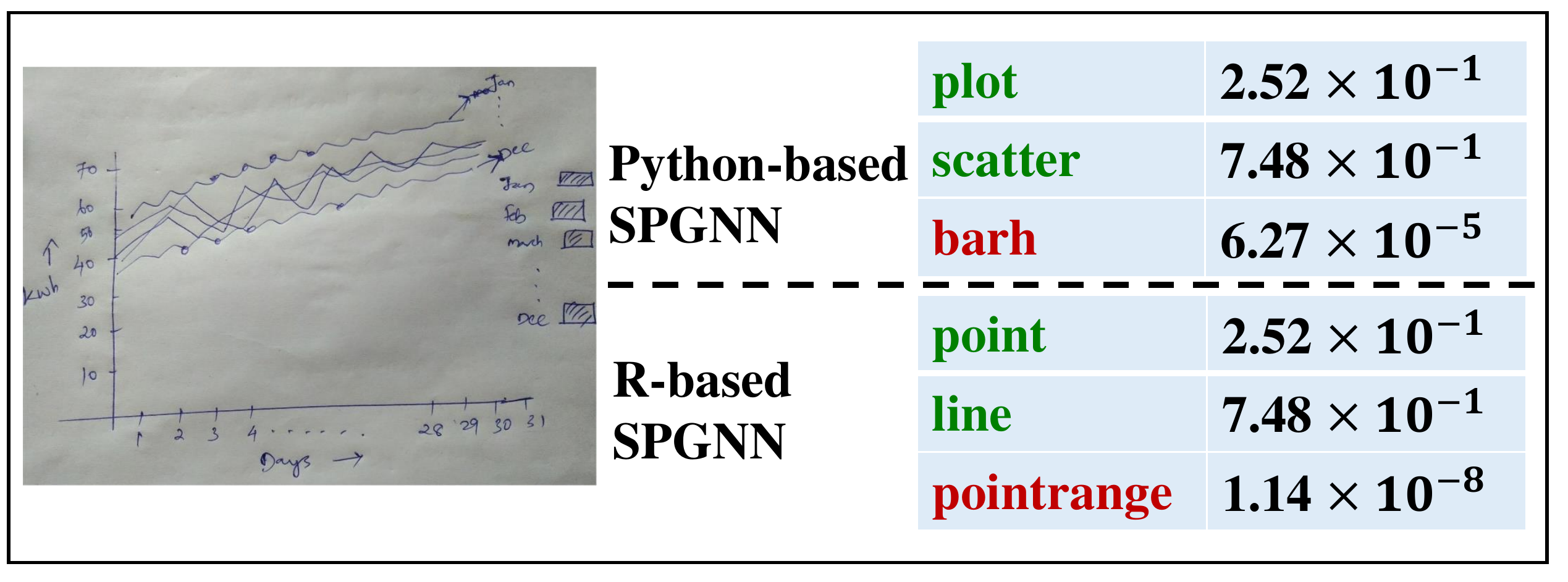}
   \label{eg2}
 }
\vspace{-0.2cm}
\caption{Several Plot2API Examples. The top-3 APIs recommended by the different models via giving the plot. The green ones are the correct APIs while the red ones are the wrong API.}\label{egs}		
\vspace{-0.4cm}
\end{figure}

\begin{figure}[t]
\centering
\includegraphics[width=0.86\columnwidth]{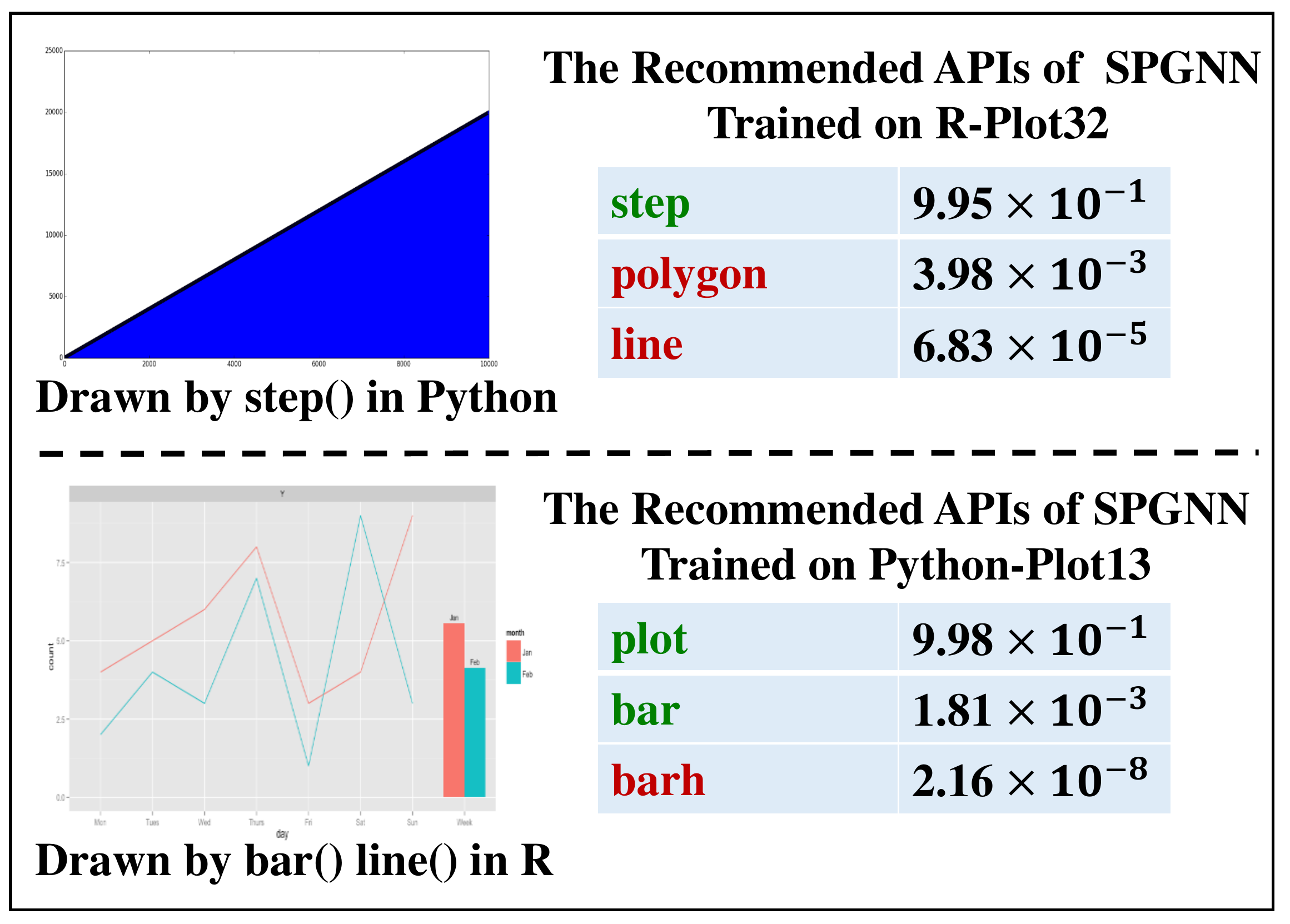}
\vspace{-0.4cm}
\caption{Two examples of cross-language API conversion. The green ones are the correct APIs while the red ones are the wrong API.}
\label{eg3}
\vspace{-0.6cm}
\end{figure}

To validate the effectiveness of our method, we visualize several practical applications for demonstrating the utility of our model in reality. Figures~\ref{egs} and \ref{eg3} show plot-based API recommendation and plot-based cross-language API conversion user scenarios respectively.

Consider the sample of the ``Breakdown of building types'' shown in the first case in Figure~\ref{eg1}, we suppose that Peter, a developer with little experience, needs to do a similar project to show the newly breakdown of building types. Therefore, the task of Peter is to plot a similar figure to demonstrate the data. If Peter knows which API can draw the figure, he can report the presentation successfully. To solve the drawing problem, he can use Plot2API model for API recommendation. In this step, the only thing he needs to do is to input the graphic in Figure~\ref{eg1} (such graphics may be just downloaded from web or acquired from other documents) to our tool, and then the tool will recommend the relevant APIs. There is another circumstance that Peter does not have a similar figure. So he has to draw a figure manually by himself. Then, he can do the same workflow with our tool to acquire the recommended APIs just based on this hand-drawn figure as the case shown in Figure~\ref{eg2}.

In agile development, some junior developers may not have broad knowledge of different programming languages and there are many software projects have similar modules or functions that can be referenced. In such a manner, the developers expect to use the output plots in some old projects developed with the familiar languages as the cues to obtain the APIs in other language which can draw the similar figures directly to accelerate the development process. Our method can support such a plot-based cross-language API recommendation scheme. Figure~\ref{eg3} shows two successful examples in this scheme. The first case is a R-Plot32 trained SPGNN gives the reasonable API recommendation for a plot drawn by Python language while the second one is a Python-Plot13 trained SPGNN recommends the correct APIs for a plot drawn by R language.

\subsection{Threats To Validity }
Since our tool is limited to Python and R programming languages, our techniques may not generalize for other programming languages. However, if the features of figures drawn from other programming languages are similar to R or Python, our tool may still work at these languages. With regard to the application to the other programming languages, we believe that our method can still success if the training data is sufficient.

The other issue is that the performance of API recommendation in some APIs of Python and R is not very well. This is due to the insufficient training data and the extremely similar characteristics of different APIs in visual appearance. The increased training samples of these APIs can address this issue, since the abundant data can facilitate SPGNN to learn more visual knowledge to better distinguish the APIs particularly the similar APIs with each other.

We only pick up some same named APIs between Python and R programming languages for validating the cross-language API recommendation due to the lack of ground truth of automatic evaluation. The manual verification will be conducted to make a more comprehensive verification in the future.

\section{Related Work}
\textbf{API Recommendation:} There are a lot of impressive works in API recommendation~\cite{thung2013automatic, raghothaman2016swim, ye2016word, gu2016deep, xu2019mulapi, cai2019biker}. The most common way for API recommendation is to rank APIs via using the similarity between the natural language query and the API description, and then recommend the APIs according to the ranks. For example, Rahman et al.~\cite{RACK} offered a recommendation of the relevant API list by using keyword-API mapping from the crowdsourced knowledge of Stack Overflow. Huang et al.~\cite{huang2018api} proposed BIKER to tackle the lexical gap and knowledge gap, so that BIKER could automatically recommend relevant APIs for a programming task described in natural language. Besides the natural language query, source code is also an important cue for API recommendation, several researchers work in this direction. McMillan et al.~\cite{mcmillan2011portfolio} proposed Portfolio to find highly relevant APIs and projects from a large archive of C/C++ source code. Chan et al. \cite{chan2012searching} improved the Portfolio by employing further sophisticated graph-mining and textual similarity techniques. A graph-based statistical language model named GraLan was proposed to develop an API suggestion engine via computing the probability of usage graphs which were learned from a corpus of source code to compute the probability of usage~\cite{nguyen2015graph}.

In conclusion, the existing API recommendation works are quite different from us. They used the natural language query or source code as cues for API recommendation task in these works, which is essentially a text to text pure Natural Language Processing (NLP) task while Plot2API is an image to text cross-model machine learning task.


\textbf{Visual Semantic Embedding:} Semantics are widely used in many neural network models for boosting the visual recognition or classification~\cite{lu202012, chen2019multi, he2016deep}, since the visual recognition models are often limited by the increasing difficulty of obtaining sufficient training data in the form of labeled images as the number of object categories grows~\cite{frome2013devise}. For example, Wang et al.~\cite{wang2016cnn} utilized recurrent neural networks(RNNs) to address the label dependencies in an image. By combining CNNs, the proposed CNN-RNN model learned both the semantic redundancy and the co-occurrence dependency in an end-to-end way. To improve multi-label image classification, Zhu et al.~\cite{zhu2017learning} proposed a unified deep neural network to capture both semantic and spatial relations of these multiple labels based on weighted attention maps. A generic structured model proposed in~\cite{hu2016learning} employed a stacked label prediction neural network, capturing both inter-level and intra-level label semantics to improve image classification performance.

\textbf{Multi-task Learning:} Multi-task learning is a popular machine technique. It aims at developing an integrated model, which can tackle multiple relevant tasks simultaneously, to exploit the complementary information among tasks for further benefiting the solution of each task~\cite{caruana1997multitask}. The multi-task learning works often enjoy a better generalization ability than the single-task learning method, and have already been successfully applied to many domains such as computer vision \cite{bragman2019stochastic, misra2016cross,strezoski2019many}, medical image analysis~\cite{wu2018joint,hussein2017risk,khosravan2018semi}, and natural language processing~\cite{collobert2008unified,liu2019multi,raffel2019exploring}, and so on. For example, Sanh et al~\cite{sanh2019hierarchical} proposed a hierarchically supervised multi-task learning model focused on a set of semantic tasks, such as entity recognition and entity mention detection.
Liu et al.~\cite{liu2020multi} presented a multi-task framework to guide the generation of TIR-specific discriminative features for distinguishing the TIR objects belonging to different classes and fine-grained correlation features for TIR tracking. Lu et al.~\cite{lu202012} studied the correlation between vision-and-language tasks for large-scale, multi-modal, multi-task learning, which shown significant gains over independent task training. Inspired by these successes, our method intends to introduce the extra semantic parsing task to boost the performance of API recommendation.

\section{Conclusions and Future Work}
In this paper, we cast a novel and meaningful software engineering task named Plot2API. To address such an issue, a deep multi-task learning method named Semantic Parsing Guided Neural Network  (SPGNN) is presented. SPGNN introduces the plot-based semantic parsing to the EfficientNet for pairing the semantic parsing of plots with the plot-based API-recommendation. Then the semantics of APIs can be exploited via the semantic parsing module for boosting the plot-based API recommendation. Three new Plot2API datasets named Python-Plot13, R-Plot32 and R-Plot14 are released for evaluation. The experimental results demonstrate the superiority over other deep learning baselines for Plot2API with a significant advantage and validate the effectiveness of our method in some application contexts of software engineering.


\section*{Acknowledge}
This work was in part supported by the National Natural Science Foundations of China (NO. 61772093 and 62002034), the Fundamental Research Funds for the Central Universities (NO. 2019CDCGRJ314, 2019CDYGYB014 and 2020CDCGRJ072).

\balance
\bibliographystyle{IEEEtran}
\bibliography{ref}

\begin{thebibliography}{10}
\providecommand{\url}[1]{#1}
\csname url@samestyle\endcsname
\providecommand{\newblock}{\relax}
\providecommand{\bibinfo}[2]{#2}
\providecommand{\BIBentrySTDinterwordspacing}{\spaceskip=0pt\relax}
\providecommand{\BIBentryALTinterwordstretchfactor}{4}
\providecommand{\BIBentryALTinterwordspacing}{\spaceskip=\fontdimen2\font plus
\BIBentryALTinterwordstretchfactor\fontdimen3\font minus
  \fontdimen4\font\relax}
\providecommand{\BIBforeignlanguage}[2]{{%
\expandafter\ifx\csname l@#1\endcsname\relax
\typeout{** WARNING: IEEEtran.bst: No hyphenation pattern has been}%
\typeout{** loaded for the language `#1'. Using the pattern for}%
\typeout{** the default language instead.}%
\else
\language=\csname l@#1\endcsname
\fi
#2}}
\providecommand{\BIBdecl}{\relax}
\BIBdecl

\bibitem{huang2018api}
Q.~Huang, X.~Xia, Z.~Xing, D.~Lo, and X.~Wang, ``Api method recommendation
  without worrying about the task-api knowledge gap,'' in \emph{2018 33rd
  IEEE/ACM International Conference on Automated Software Engineering
  (ASE)}.\hskip 1em plus 0.5em minus 0.4em\relax IEEE, 2018, pp. 293--304.

\bibitem{mcmillan2011portfolio}
C.~McMillan, M.~Grechanik, D.~Poshyvanyk, Q.~Xie, and C.~Fu, ``Portfolio:
  finding relevant functions and their usage,'' in \emph{Proceedings of the
  33rd International Conference on Software Engineering}, 2011, pp. 111--120.

\bibitem{campbell2017nlp2code}
B.~A. Campbell and C.~Treude, ``Nlp2code: Code snippet content assist via
  natural language tasks,'' in \emph{2017 IEEE International Conference on
  Software Maintenance and Evolution (ICSME)}.\hskip 1em plus 0.5em minus
  0.4em\relax IEEE, 2017, pp. 628--632.

\bibitem{allamanis2015bimodal}
M.~Allamanis, D.~Tarlow, A.~Gordon, and Y.~Wei, ``Bimodal modelling of source
  code and natural language,'' in \emph{International conference on machine
  learning}, 2015, pp. 2123--2132.

\bibitem{gvero2015interactive}
T.~Gvero and V.~Kuncak, ``Interactive synthesis using free-form queries,'' in
  \emph{2015 IEEE/ACM 37th IEEE International Conference on Software
  Engineering}, vol.~2.\hskip 1em plus 0.5em minus 0.4em\relax IEEE, 2015, pp.
  689--692.

\bibitem{nguyen2018statistical}
A.~Nguyen, P.~Rigby, T.~Nguyen, D.~Palani, M.~Karanfil, and T.~Nguyen,
  ``Statistical translation of english texts to api code templates,'' in
  \emph{2018 IEEE International Conference on Software Maintenance and
  Evolution (ICSME)}.\hskip 1em plus 0.5em minus 0.4em\relax IEEE, 2018, pp.
  194--205.

\bibitem{tan2019efficientnet}
M.~Tan and Q.~Le, ``Efficientnet: Rethinking model scaling for convolutional
  neural networks,'' in \emph{International Conference on Machine Learning},
  2019, pp. 6105--6114.

\bibitem{GoogLeNet}
C.~Szegedy, W.~Liu, Y.~Jia, P.~Sermanet, S.~Reed, D.~Anguelov, D.~Erhan,
  V.~Vanhoucke, and A.~Rabinovich, ``Going deeper with convolutions,'' in
  \emph{Proceedings of the IEEE conference on computer vision and pattern
  recognition}, 2015, pp. 1--9.

\bibitem{simonyan2014VGG}
K.~Simonyan and A.~Zisserman, ``Very deep convolutional networks for
  large-scale image recognition,'' \emph{arXiv preprint arXiv:1409.1556}, 2014.

\bibitem{mlgcn}
Z.-M. Chen, X.-S. Wei, P.~Wang, and Y.~Guo, ``Multi-label image recognition
  with graph convolutional networks,'' in \emph{Proceedings of the IEEE
  Conference on Computer Vision and Pattern Recognition}, 2019, pp. 5177--5186.

\bibitem{cnnplp}
L.~Mou, G.~Li, L.~Zhang, T.~Wang, and Z.~Jin, ``Convolutional neural networks
  over tree structures for programming language processing,'' in
  \emph{Proceedings of the Thirtieth AAAI Conference on Artificial
  Intelligence}, 2016, pp. 1287--1293.

\bibitem{rasiwasia2010new}
N.~Rasiwasia, J.~Costa~Pereira, E.~Coviello, G.~Doyle, G.~R. Lanckriet,
  R.~Levy, and N.~Vasconcelos, ``A new approach to cross-modal multimedia
  retrieval,'' in \emph{Proceedings of the 18th ACM international conference on
  Multimedia}, 2010, pp. 251--260.

\bibitem{mikolov2013distributed}
T.~Mikolov, I.~Sutskever, K.~Chen, G.~S. Corrado, and J.~Dean, ``Distributed
  representations of words and phrases and their compositionality,'' in
  \emph{Advances in neural information processing systems}, 2013, pp.
  3111--3119.

\bibitem{RN}
F.~Sung, Y.~Yang, L.~Zhang, T.~Xiang, P.~H. Torr, and T.~M. Hospedales,
  ``Learning to compare: Relation network for few-shot learning,'' in
  \emph{Proceedings of the IEEE Conference on Computer Vision and Pattern
  Recognition}, 2018, pp. 1199--1208.

\bibitem{zhong2020random}
Z.~Zhong, L.~Zheng, G.~Kang, S.~Li, and Y.~Yang, ``Random erasing data
  augmentation.'' in \emph{AAAI}, 2020, pp. 13\,001--13\,008.

\bibitem{deng2009imagenet}
J.~Deng, W.~Dong, R.~Socher, L.-J. Li, K.~Li, and L.~Fei-Fei, ``Imagenet: A
  large-scale hierarchical image database,'' in \emph{2009 IEEE conference on
  computer vision and pattern recognition}.\hskip 1em plus 0.5em minus
  0.4em\relax Ieee, 2009, pp. 248--255.

\bibitem{tan2019mnasnet}
M.~Tan, B.~Chen, R.~Pang, V.~Vasudevan, M.~Sandler, A.~Howard, and Q.~V. Le,
  ``Mnasnet: Platform-aware neural architecture search for mobile,'' in
  \emph{Proceedings of the IEEE Conference on Computer Vision and Pattern
  Recognition}, 2019, pp. 2820--2828.

\bibitem{sandler2018mobilenetv2}
M.~Sandler, A.~Howard, M.~Zhu, A.~Zhmoginov, and L.-C. Chen, ``Mobilenetv2:
  Inverted residuals and linear bottlenecks,'' in \emph{Proceedings of the IEEE
  conference on computer vision and pattern recognition}, 2018, pp. 4510--4520.

\bibitem{kingma2014adam}
D.~P. Kingma and J.~Ba, ``Adam: A method for stochastic optimization,''
  \emph{arXiv preprint arXiv:1412.6980}, 2014.

\bibitem{he2016Resnet}
K.~He, X.~Zhang, S.~Ren, and J.~Sun, ``Deep residual learning for image
  recognition,'' in \emph{Proceedings of the IEEE conference on computer vision
  and pattern recognition}, 2016, pp. 770--778.

\bibitem{thung2013automatic}
F.~Thung, S.~Wang, D.~Lo, and J.~Lawall, ``Automatic recommendation of api
  methods from feature requests,'' in \emph{Proceedings of International
  Conference on Automated Software Engineering}, 2013, pp. 290--300.

\bibitem{raghothaman2016swim}
M.~Raghothaman, Y.~Wei, and Y.~Hamadi, ``Swim: Synthesizing what i mean-code
  search and idiomatic snippet synthesis,'' in \emph{Proceedings of
  International Conference on Software Engineering}, 2016, pp. 357--367.

\bibitem{ye2016word}
X.~Ye, H.~Shen, X.~Ma, R.~Bunescu, and C.~Liu, ``From word embeddings to
  document similarities for improved information retrieval in software
  engineering,'' in \emph{Proceedings of International Conference on Software
  Engineering}, 2016, pp. 404--415.

\bibitem{gu2016deep}
X.~Gu, H.~Zhang, D.~Zhang, and S.~Kim, ``Deep api learning,'' in
  \emph{Proceedings of ACM SIGSOFT International Symposium on Foundations of
  Software Engineering}, 2016, pp. 631--642.

\bibitem{xu2019mulapi}
C.~Xu, B.~Min, X.~Sun, J.~Hu, B.~Li, and Y.~Duan, ``Mulapi: A tool for api
  method and usage location recommendation,'' in \emph{Proceedings of
  International Conference on Software Engineering: Companion Proceedings},
  2019, pp. 119--122.

\bibitem{cai2019biker}
L.~Cai, H.~Wang, Q.~Huang, X.~Xia, Z.~Xing, and D.~Lo, ``Biker: a tool for
  bi-information source based api method recommendation,'' in \emph{Proceedings
  of ACM Joint Meeting on European Software Engineering Conference and
  Symposium on the Foundations of Software Engineering}, 2019, pp. 1075--1079.

\bibitem{RACK}
M.~M. Rahman, C.~K. Roy, and D.~Lo, ``Rack: Automatic api recommendation using
  crowdsourced knowledge,'' in \emph{2016 IEEE 23rd International Conference on
  Software Analysis, Evolution, and Reengineering (SANER)}, vol.~1.\hskip 1em
  plus 0.5em minus 0.4em\relax IEEE, 2016, pp. 349--359.

\bibitem{chan2012searching}
W.-K. Chan, H.~Cheng, and D.~Lo, ``Searching connected api subgraph via text
  phrases,'' in \emph{Proceedings of the ACM SIGSOFT 20th International
  Symposium on the Foundations of Software Engineering}, 2012, pp. 1--11.

\bibitem{nguyen2015graph}
A.~T. Nguyen and T.~N. Nguyen, ``Graph-based statistical language model for
  code,'' in \emph{2015 IEEE/ACM 37th IEEE International Conference on Software
  Engineering}, vol.~1.\hskip 1em plus 0.5em minus 0.4em\relax IEEE, 2015, pp.
  858--868.

\bibitem{lu202012}
J.~Lu, V.~Goswami, M.~Rohrbach, D.~Parikh, and S.~Lee, ``12-in-1: Multi-task
  vision and language representation learning,'' in \emph{Proceedings of the
  IEEE/CVF Conference on Computer Vision and Pattern Recognition}, 2020, pp.
  10\,437--10\,446.

\bibitem{chen2019multi}
Z.-M. Chen, X.-S. Wei, P.~Wang, and Y.~Guo, ``Multi-label image recognition
  with graph convolutional networks,'' in \emph{Proceedings of the IEEE
  Conference on Computer Vision and Pattern Recognition}, 2019, pp. 5177--5186.

\bibitem{he2016deep}
K.~He, X.~Zhang, S.~Ren, and J.~Sun, ``Deep residual learning for image
  recognition,'' in \emph{Proceedings of the IEEE conference on computer vision
  and pattern recognition}, 2016, pp. 770--778.

\bibitem{frome2013devise}
A.~Frome, G.~S. Corrado, J.~Shlens, S.~Bengio, J.~Dean, M.~Ranzato, and
  T.~Mikolov, ``Devise: A deep visual-semantic embedding model,'' in
  \emph{Advances in neural information processing systems}, 2013, pp.
  2121--2129.

\bibitem{wang2016cnn}
J.~Wang, Y.~Yang, J.~Mao, Z.~Huang, C.~Huang, and W.~Xu, ``Cnn-rnn: A unified
  framework for multi-label image classification,'' in \emph{Proceedings of the
  IEEE conference on computer vision and pattern recognition}, 2016, pp.
  2285--2294.

\bibitem{zhu2017learning}
F.~Zhu, H.~Li, W.~Ouyang, N.~Yu, and X.~Wang, ``Learning spatial regularization
  with image-level supervisions for multi-label image classification,'' in
  \emph{Proceedings of the IEEE Conference on Computer Vision and Pattern
  Recognition}, 2017, pp. 5513--5522.

\bibitem{hu2016learning}
H.~Hu, G.-T. Zhou, Z.~Deng, Z.~Liao, and G.~Mori, ``Learning structured
  inference neural networks with label relations,'' in \emph{Proceedings of the
  IEEE Conference on Computer Vision and Pattern Recognition}, 2016, pp.
  2960--2968.

\bibitem{caruana1997multitask}
R.~Caruana, ``Multitask learning,'' \emph{Machine learning}, vol.~28, no.~1,
  pp. 41--75, 1997.

\bibitem{bragman2019stochastic}
F.~J. Bragman, R.~Tanno, S.~Ourselin, D.~C. Alexander, and J.~Cardoso,
  ``Stochastic filter groups for multi-task cnns: Learning specialist and
  generalist convolution kernels,'' in \emph{Proceedings of the IEEE
  International Conference on Computer Vision}, 2019, pp. 1385--1394.

\bibitem{misra2016cross}
I.~Misra, A.~Shrivastava, A.~Gupta, and M.~Hebert, ``Cross-stitch networks for
  multi-task learning,'' in \emph{Proceedings of the IEEE Conference on
  Computer Vision and Pattern Recognition}, 2016, pp. 3994--4003.

\bibitem{strezoski2019many}
G.~Strezoski, N.~v. Noord, and M.~Worring, ``Many task learning with task
  routing,'' in \emph{Proceedings of the IEEE International Conference on
  Computer Vision}, 2019, pp. 1375--1384.

\bibitem{wu2018joint}
B.~Wu, Z.~Zhou, J.~Wang, and Y.~Wang, ``Joint learning for pulmonary nodule
  segmentation, attributes and malignancy prediction,'' in \emph{2018 IEEE 15th
  International Symposium on Biomedical Imaging (ISBI 2018)}.\hskip 1em plus
  0.5em minus 0.4em\relax IEEE, 2018, pp. 1109--1113.

\bibitem{hussein2017risk}
S.~Hussein, K.~Cao, Q.~Song, and U.~Bagci, ``Risk stratification of lung
  nodules using 3d cnn-based multi-task learning,'' in \emph{International
  conference on information processing in medical imaging}.\hskip 1em plus
  0.5em minus 0.4em\relax Springer, 2017, pp. 249--260.

\bibitem{khosravan2018semi}
N.~Khosravan and U.~Bagci, ``Semi-supervised multi-task learning for lung
  cancer diagnosis,'' in \emph{2018 40th Annual International Conference of the
  IEEE Engineering in Medicine and Biology Society (EMBC)}.\hskip 1em plus
  0.5em minus 0.4em\relax IEEE, 2018, pp. 710--713.

\bibitem{collobert2008unified}
R.~Collobert and J.~Weston, ``A unified architecture for natural language
  processing: Deep neural networks with multitask learning,'' in
  \emph{Proceedings of the 25th international conference on Machine learning},
  2008, pp. 160--167.

\bibitem{liu2019multi}
X.~Liu, P.~He, W.~Chen, and J.~Gao, ``Multi-task deep neural networks for
  natural language understanding,'' in \emph{Proceedings of the 57th Annual
  Meeting of the Association for Computational Linguistics}, 2019, pp.
  4487--4496.

\bibitem{raffel2019exploring}
C.~Raffel, N.~Shazeer, A.~Roberts, K.~Lee, S.~Narang, M.~Matena, Y.~Zhou,
  W.~Li, and P.~J. Liu, ``Exploring the limits of transfer learning with a
  unified text-to-text transformer,'' \emph{arXiv preprint arXiv:1910.10683},
  2019.

\bibitem{sanh2019hierarchical}
V.~Sanh, T.~Wolf, and S.~Ruder, ``A hierarchical multi-task approach for
  learning embeddings from semantic tasks,'' in \emph{Proceedings of the AAAI
  Conference on Artificial Intelligence}, vol.~33, 2019, pp. 6949--6956.

\bibitem{liu2020multi}
Q.~Liu, X.~Li, Z.~He, N.~Fan, D.~Y. 0002, W.~Liu, and Y.~Liang, ``Multi-task
  driven feature models for thermal infrared tracking.'' in \emph{AAAI}, 2020,
  pp. 11\,604--11\,611.

\end{thebibliography}
\end{document}